\documentclass{article}
\usepackage{bookstyle}
\usepackage{graphicx,amsfonts,amssymb,verbatim}
\usepackage{cite}

\begin{document}

\setlength\arraycolsep{2pt}

\newcommand{\be}{\begin{equation}}
\newcommand{\ee}{\end{equation}}
\newcommand{\bea}{\begin{eqnarray}}
\newcommand{\eea}{\end{eqnarray}}
\newcommand{\mt}{\tilde{m}}

\newcommand{\eref}[1]{(\ref{#1})}
\newcommand{\R}[1]{\mathbb{R}^{#1}}
\newcommand{\vev}[1]{\langle#1\rangle}
\def\NN{{\cal N}}
\def\none{$\NN=1$}
\def\RR{{\bf R}}
\def\tr{{\rm tr}}
\def\Tr{{\rm Tr}}
\def\ZZ{{\mathbb Z}}

\author{Marcus K. Benna and Igor R. Klebanov}

\address{Joseph Henry Laboratories and Princeton Center for Theoretical Physics, \\
Princeton University, Princeton, NJ 08544, USA
%\\E-mail: klebanov@princeton.edu
}

\title{Gauge--String Dualities and Some Applications\quad}

\photo{}

\frontmatter
\maketitle    
\mainmatter

\abstract{
The first part of these lectures contains an introductory review of the AdS/CFT duality and of its
tests. Applications to thermal gauge theory are also discussed briefly. The second part is devoted
to a review of gauge-string dualities based on various warped conifold backgrounds, and to their
cosmological applications.
}

\section{Introduction}

String theory 
is well known to be the leading prospect for 
quantizing gravity and
unifying it with other interactions. 
One may also take a broader view of string
theory as a description of string-like excitations that 
arise in many different
physical systems, such as the superconducting 
flux tubes or
the chromo-electric flux tubes in non-Abelian gauge theories.
From the point of view of quantum field theories describing the physical systems 
where these string-like objects arise, they 
are ``emergent''
rather than fundamental. However, thanks to the AdS/CFT correspondence
\cite{jthroat,US,EW} and its extensions, we now know that at least some field theories have dual formulations
in terms of string theories in curved backgrounds. In these examples, the strings
that are ``emergent'' from the field theory point of view are dual to fundamental
or D-strings in the string theoretic approach. 
Besides being of great theoretical
interest, such dualities are becoming a useful tool
for studying strongly coupled gauge theories. 
These ideas also have far-reaching implications for
building connections between string theory and the real world.

These notes, based on five lectures delivered by I.R.K. at Les Houches in July 2007, are
not meant to be a comprehensive review of what has become a vast field. 
Instead, they aim to present a particular path through it, which begins with old and well-known concepts,
and eventually leads to some recent developments. The first part of these lectures is based on
an earlier brief review \cite{qcdstring}.
It begins with a bit of history and basic facts about string theory and
its connections with strong interactions. Comparisons
of stacks of Dirichlet branes
with curved backgrounds produced by them are used to motivate the AdS/CFT
correspondence between superconformal gauge theory and
string theory on a product of Anti-de Sitter space and a compact manifold.
The ensuing duality between semi-classical spinning strings
and gauge theory operators carrying large charges is briefly reviewed. 
We go on to describe recent tests of the AdS/CFT correspondence 
using the Wilson loop cusp anomaly as a function of the
coupling, which also enters dimensions of high-spin operators.
Strongly coupled thermal SYM theory is explored via
a black hole in 5-dimensional AdS space, which 
leads to explicit results for its
entropy and shear viscosity.

The second part of these lectures (sections 7--11)
focuses on the gauge-string dualities that arise from
studying D-branes on the conifold geometry. The $AdS_5\times T^{1,1}$ background appears
as the type IIB dual of an $SU(N)\times SU(N)$ superconformal
gauge theory with bi-fundamental fields. A warped resolved conifold is then presented
as a description of holographic RG flow from this theory to the ${\cal N}=4$
supersymmetric gauge theory produced by giving a classical value to one of the bi-fundamentals. 
The warped deformed conifold is instead the dual of the cascading $SU(kM)\times 
SU((k+1)M)$ gauge theory which confines in the infrared. Some features of the
bound state spectrum are discussed, and the supergravity dual of the baryonic branch,
the resolved warped deformed conifold, is reviewed. We end with a brief update on possible 
cosmological applications of these backgrounds.

\section{Strings and QCD}\label{subsec:prod}

String theory was born out of attempts to understand
    the strong interactions.
    Empirical evidence for a string-like structure of hadrons 
comes from arranging mesons and baryons into approximately
linear Regge trajectories. Studies of  
$\pi N$ scattering prompted
    Dolen, Horn and Schmid\cite{Dolen} to make a duality 
conjecture stating that 
the sum over s-channel exchanges 
equals the sum over t-channel ones. This posed the problem of finding
the analytic form of such dual amplitudes.
Veneziano\cite{Veneziano} found the first, and very simple, expression for a
 manifestly dual 4-point amplitude:
\be 
A(s,t) \sim {\Gamma (-\alpha(s))
\Gamma (-\alpha(t))\over
\Gamma (-\alpha(s) - \alpha(t)) }
\ee
with an exactly linear Regge trajectory
$ \alpha(s) = \alpha(0) + \alpha' s$.
Soon after, its open string interpretation was proposed
\cite{Nambu,Nielsen,Susskind} (for detailed reviews of these early developments,
see \cite{DiVecchia}).
In the early 70's this led to an explosion of interest in string
theory as a description of strongly interacting particles.
The basic idea is to think of a meson as an open string
with a quark at one end-point and an anti-quark at the other. Then various
meson states arise as rotational and vibrational excitations of such an open string.
The splitting of such a string describes the decay of a meson into two mesons.

The string world sheet dynamics is governed by the Nambu-Goto area action
\be S_{\rm NG}= -T \int d\sigma d\tau
\sqrt{-{\rm det}\ \partial_a X^\mu \partial_b X_\mu}
\ ,
\ee
where the indices $a,b$ take two values ranging over the
$\sigma$ and $\tau$ directions on the world sheet.
The string tension is related to the Regge slope through
$ T^{-1}= 2\pi\alpha' $.
The quantum consistency of the Veneziano model requires  that the Regge 
intercept is $\alpha(0)=1$,
so that the spin 1 state is massless but the spin 0 is a tachyon. 
But the $\rho$ meson is certainly not massless, and the presence of a 
tachyon in the spectrum indicates an instability. This is how
the string theory of strong interactions started
to run into problems.

Calculation of the string zero-point energy gives
$\alpha(0) = {(d-2)/ 24} $.
Hence the model has to be defined in 26 space-time  dimensions.
Consistent supersymmetric string theories were discovered in 10 
dimensions, but their relation to strong interactions in 4 dimensions was 
initially completely unclear.
Most importantly, the Asymptotic Freedom of strong interactions was 
discovered\cite{GWP}, 
singling out Quantum Chromodynamics (QCD) as the exact 
field theory of strong interactions.
At this point
most physicists gave up on strings as a description of strong interactions. 
Instead, since 
the graviton appears naturally in the closed string 
spectrum,       
string theory emerged as the leading hope for unifying 
quantum gravity with other forces\cite{Scherk,Yoneya}.

Now that we know that a non-Abelian gauge theory is an exact
description of strong interactions, is there any room left for
string theory in this field? Luckily, the answer is positive.
At short distances, much smaller than 1 fermi, 
the quark anti-quark potential is approximately Coulombic due to 
the Asymptotic Freedom.
At large distances the potential should be linear  
due to formation of a confining flux tube\cite{Wilson}.
When these tubes are much longer than their thickness, one can hope to
describe them, at least
approximately, by semi-classical Nambu strings\cite{Nambunew}.
This appears to explain the existence of approximately linear
Regge trajectories. For the leading trajectory,
a linear relation between angular momentum and mass-squared 
\be J =\alpha' m^2 + \alpha (0)
\ ,
\ee
is provided 
by a semi-classical spinning relativistic string with massless quark 
and anti-quark at its endpoints. In case of baryons, one finds a di-quark instead
of an anti-quark at one of the endpoints.
A semi-classical string approach to the QCD flux tubes 
is widely used, for example, 
in jet hadronization algorithms based on the 
Lund String Model\cite{Andersson}. 

Semi-classical quantization around a long straight 
Nambu string predicts the quark anti-quark 
potential\cite{Luscher:1980fr}
\be
V(r) = Tr + \mu +{\gamma\over r} + O(1/r^2)
\ .\ee
The coefficient $\gamma$ 
of the universal L\" uscher term depends only on the 
space-time dimension $d$ and is proportional to the Regge intercept:
$\gamma=-\pi (d-2)/24$.
Lattice calculations of the force vs. distance for
probe quarks and anti-quarks\cite{Luscher} produce good 
agreement with this value in $d=3$ and $d=4$ 
for $r>0.7$ fm.
Thus, long QCD strings appear to be well described by the Nambu-Goto
area action. But quantization of short, highly quantum QCD strings,
that could lead to a calculation of light meson and glueball spectra,
is a much harder problem.

The connection of gauge theory with string theory is strengthened by
`t Hooft's generalization of QCD from 3 colors 
($SU(3)$ gauge group) to $N$ colors ($SU(N)$ gauge group)\cite{GT}.
The idea is to make $N$ 
large, while keeping the `t Hooft coupling 
$\lambda = g_{\rm YM}^2 N$ fixed.
In this limit each Feynman graph carries
a topological factor $N^\chi$, where $\chi$ is the Euler
characteristic of the graph.
Thus, the sum over graphs of a given topology can perhaps be thought of
as a sum over world sheets of a hypothetical ``QCD string.''
Since the spheres (string tree diagrams)
are weighted by $N^2$, the tori (string one-loop diagrams)
by $N^0$ etc., we find that the closed string coupling constant is of order
$N^{-1}$. Thus, the advantage of taking $N$ to be large is that we find
a weakly coupled string theory. 
In the large $N$ limit the gauge theory 
simplifies in that 
only the planar diagrams contribute. But 
directly summing even this subclass of diagrams seems to be an
impossible task.
From the dual QCD string point of view, it
is not clear how to describe
this string theory in elementary terms.

Because of the difficulty of these problems,
between the late 70's and the
mid-90's many theorists gave up hope of finding an exact gauge-string duality.
One notable exception is Polyakov who in 1981 proposed that 
the string theory dual to a 
4-d gauge theory should have a 5-th hidden dimension\cite{Polyakov}. 
In later work\cite{Polyakovnew} he refined this proposal, 
suggesting that the 5-d metric must
be ``warped.''

\section{The Geometry of Dirichlet Branes}

In the mid-90's Dirichlet branes,
or D-branes for short,
 brought string theory back to 
gauge theory.
D-branes are soliton-like ``membranes'' of various internal dimensionalities
contained in theories of closed superstrings\cite{Polch}.
A Dirichlet $p$-brane (or D$p$-brane) is a $p+1$ dimensional hyperplane
in $9+1$ dimensional space-time where strings are allowed to end.
A D-brane is much like a topological defect: upon 
touching a D-brane, a closed string
can open up and turn into an open string whose
ends are free to move along the D-brane. For the end-points of such a string
the $p+1$ longitudinal coordinates satisfy the conventional free (Neumann)
boundary conditions, while the $9-p$ coordinates transverse to
the D$p$-brane have the fixed (Dirichlet) boundary conditions; hence
the origin of the term ``Dirichlet brane.'' 
In a seminal paper\cite{Polch} Polchinski
showed that a D$p$-brane 
preserves
$1/2$ of the bulk supersymmetries and carries an elementary unit
of charge with respect to the $p+1$ form gauge potential from the
Ramond-Ramond sector of type II superstring. 

For our purposes, the most important property of D-branes 
is that they realize
gauge theories on their world volume. The massless spectrum of open strings
living on a D$p$-brane is that of a maximally supersymmetric $U(1)$
gauge theory in $p+1$ dimensions. The $9-p$ massless
scalar fields present in this supermultiplet are the expected Goldstone 
modes associated with the transverse oscillations of the D$p$-brane,
while the photons and fermions
provide the unique supersymmetric
completion.
If we consider $N$ parallel D-branes,
then there are $N^2$ different species of open strings because they can
begin and end on any of the D-branes. 
$N^2$ is the dimension of the adjoint representation of $U(N)$,
and indeed we find the maximally supersymmetric $U(N)$ 
gauge theory in this setting.

The relative separations of the D$p$-branes in the $9-p$ transverse
dimensions are determined by the expectation values of the scalar fields.
We will be 
interested in the case where all scalar expectation
values vanish, so that the $N$ D$p$-branes are stacked on top of each other.
If $N$ is large, then this stack is a heavy object embedded into a theory
of closed strings which contains gravity. Naturally, this macroscopic
object will curve space: it may be described by some classical metric
and other background fields.
Thus, we have two very different descriptions of the stack of D$p$-branes:
one in terms of the $U(N)$ supersymmetric gauge theory on its world volume,
and the other in terms of the classical
charged $p$-brane background of the type II
closed superstring theory. The relation between these two descriptions
is at the heart of the 
connections between gauge fields and strings that are the subject of
these lectures.

Parallel
D3-branes realize a $3+1$ dimensional $U(N)$ gauge theory,
which is a maximally supersymmetric ``cousin'' of QCD.
Let us compare a stack of coincident D3-branes with the 
Ramond-Ramond charged black 3-brane
classical solution whose metric assumes the form\cite{Horowitz}:
\bea
\label{metric}
   ds^2 & = &
h^{-1/2}(r)
    \left [ - f(r) (dx^0)^2 + 
\sum_{i=1}^3 
(d x^i)^2 \right] \nonumber \\
& + & h^{1/2}(r)
    \left [f^{-1} (r) dr^2 + r^2 d\Omega_{5}^2 \right ] \ ,
\eea
where 
$$   h(r)  = 1 + {L^4 \over r^4} \ , \qquad \  \  
f(r) = 1- {r_0^4\over
r^4}
\ .
$$
Here $d\Omega_5^2$ is the metric of a unit $5$ dimensional sphere,
${\bf S}^5$. 

In general, a $d$-dimensional sphere of radius $L$ 
may be defined by a constraint
\be
\sum_{i=1}^{d+1} (X^i)^2 = L^2
\ee
on $d+1$ real coordinates $X^i$. It is a positively curved 
maximally symmetric space with symmetry group $SO(d+1)$.
Similarly,
the $d$-dimensional Anti-de Sitter space, $AdS_d$,
is defined by a constraint
\be \label{embed}
(X^0)^2 + (X^d)^2 - \sum_{i=1}^{d-1} (X^i)^2 = L^2\ ,
\ee
where $L$ is its curvature radius.
$AdS_d$ is a negatively curved maximally symmetric space with
symmetry group $SO(2, d-2)$.
There exists a subspace of $AdS_d$ called the Poincar\' e wedge,
with the metric
\be \label{Poin}
ds^2 = {L^2 \over z^2} \left(dz^2 -(dx^0)^2 + \sum_{i=1}^{d-2}
(dx^i)^2\right)\ ,
\ee
where $z\in [0,\infty)$. 
In these coordinates the boundary of $AdS_d$ is at $z=0$.

The event horizon of the black 3-brane metric (\ref{metric})
is located at $r=r_0$.
In the extremal limit $r_0 \rightarrow 0$
the 3-brane metric becomes
\be
\label{geom}
ds^2 =  h^{-1/2}(r) \eta_{\mu\nu} dx^\mu dx^\nu
 + h^{1/2}(r) 
\left ( dr^2 + r^2 d\Omega_5^2 \right )\ 
\ ,
\ee
where $\eta_{\mu\nu}$ is the $3+1$ dimensional Minkowski metric.
Just like the stack of parallel, ground state D3-branes, 
the extremal solution
preserves 16 of the 32 supersymmetries present 
in the type IIB theory. Introducing $z={L^2 / r}$,
one notes that the limiting form of (\ref{geom})
as $r\rightarrow 0$ 
factorizes into the direct product of
two smooth spaces, the Poincar\' e wedge (\ref{Poin})
of $AdS_5$, 
and ${\bf S}^5$,
with equal radii of curvature $L$: 
\be \label{adsmetric}
ds^2 = {L^2 \over z^2} \left( dz^2 + \eta_{\mu\nu} dx^\mu dx^\nu \right) +
    L^2 d\Omega_5^2 \ .
\ee
The 3-brane geometry may thus be
viewed as a semi-infinite throat of radius $L$ which for
$r \gg L$ opens up into flat $9+1$ dimensional space.
Thus, for $L$ much larger than the string length scale,
$\sqrt {\alpha'}$, 
the entire 3-brane geometry has small curvatures
everywhere and is appropriately described by the supergravity
approximation to type IIB string theory.

The relation between $L$
and $\sqrt{\alpha'}$ 
may be found by equating the gravitational
tension of the
extremal 3-brane classical solution to $N$ times the tension
of a single D3-brane, and one finds
\be\label{throatrel}
L^4 = g_{\rm YM}^2 N \alpha'^2\ .
\ee 
Thus, the size of the throat in string units is $\lambda^{1/4}$.
This remarkable emergence of the `t Hooft coupling
from gravitational considerations is at the heart of the success of
the AdS/CFT correspondence. Moreover,
the requirement $L\gg \sqrt{\alpha'}$ translates into
$\lambda \gg 1$: the gravitational approach is valid when
the `t Hooft coupling is very strong and the 
perturbative field
theoretic methods are not applicable.

\section{The AdS/CFT Correspondence}

Studies of massless particle absorption by the 
3-branes \cite{absorption}
indicate that, in the low-energy limit, the 
$AdS_5 \times {\bf S}^5$ throat 
region ($r \ll L$) decouples from the asymptotically flat
large $r$ region. Similarly, the ${\cal N}=4$ supersymmetric 
$SU(N)$ gauge theory on the stack of $N$ D3-branes decouples 
in the low-energy limit from the bulk closed string theory.
Such considerations 
prompted Maldacena \cite{jthroat} to make the seminal conjecture that type IIB string 
theory on $AdS_5 \times {\bf S}^5$, of radius $L$ given in
(\ref{throatrel}), is dual to
the ${\cal N}=4$ SYM theory. The number of colors in the gauge
theory, $N$, is dual to the number of flux units of the
5-form Ramond-Ramond field strength.

It was further conjectured in \cite{US,EW} that there
exists a one-to-one map between gauge invariant operators in the CFT and
fields (or extended objects) in AdS$_5$. The dimension $\Delta$ of an
operator is determined
by the mass of the dual field $\varphi$
in AdS$_5$. For example, for scalar operators
one finds that $\Delta (\Delta-4)= m^2 L^2$. For the fields in $AdS_5$ that come from
the type IIB supergravity modes, including the Kaluza-Klein excitations on the
5-sphere, the masses are of order $1/L$. Hence, it is consistent to assume that
their operator dimensions are independent of $L$, and therefore independent of 
$\lambda$. This is due to the fact that such operators commute with
some of the supercharges and are therefore protected by supersymmetry.
Perhaps the simplest such operators are the chiral primaries which are
traceless symmetric polynomials made out of the six
scalar fields $\Phi^i$:
${\rm tr}\ \Phi^{(i_1} \ldots \Phi^{i_k)}$. These operators are dual to spherical harmonics
on ${\bf S}^5$ which mix the graviton and RR 4-form fluctuations.
Their masses are $m_k^2= k(k-4)/L^2$, where $k=2, 3, \ldots$. These masses
reproduce the operator dimensions $\Delta=k$ which are the same as in the free theory.
The situation is completely different for operators dual to the massive string modes:
$m_n^2 = 4 n/\alpha'$. In this case the AdS/CFT correspondence predicts that the operator
dimension grows at strong coupling as $2 n^{1/2} \lambda^{1/4}$. 

Precise methods
for calculating correlation functions
of various operators in a CFT using its dual formulation 
were formulated in \cite{US,EW} where
a gauge theory quantity, $W$, was identified with
a string theory quantity, $Z_{\rm string}$:
\be \label{GKPWmod}
   W[\varphi_0 (\vec x)] = Z_{\rm string} [\varphi_0 (\vec x)] 
     \ . 
\ee 
 $W$ generates the connected Euclidean
Green's functions of a gauge theory
operator $\cal O$, 
\be \label{GKPW}
   W[\varphi_0 (\vec x)] =\langle
\exp \int d^4 x \varphi_0 {\cal O} \rangle
     \ .
\ee
$Z_{\rm string}$ is the string theory path integral
calculated
as a functional of $\varphi_0$, the boundary condition on the field
$\varphi$ related to ${\cal O}$ by the AdS/CFT duality.
In the large $N$ limit the string theory becomes classical, which implies
\be
Z_{\rm string}\sim e^{- I [\phi_0 (\vec x)]}\ ,
\ee
where $I [\phi_0 (\vec x)]$
is the extremum of the classical string action calculated
as a functional of $\phi_0$.
If we are further interested in
correlation functions at very large `t Hooft coupling, then the
problem of extremizing the
classical string action reduces to solving the equations of
motion in type IIB supergravity whose form is known explicitly.

If the number of colors $N$ is sent to infinity
while $g_{\rm YM}^2 N$ is held fixed and large, then there are small
string scale corrections to the supergravity limit\cite{jthroat,US,EW}
which proceed in powers of
${\alpha' / L^2} = \lambda^{-1/2}.$
If we wish to study finite $N$, then there are also string loop
corrections in powers of
$ {\kappa^2 / L^8} \sim N^{-2}.$
As expected, taking $N$ to infinity enables us to take
the classical limit of the string theory on $AdS_5\times {\bf S}^5$.

Immediate support for the AdS/CFT correspondence comes from
symmetry considerations\cite{jthroat}. The isometry group of
$AdS_5$ is $SO(2,4)$, and this is also the conformal group in
$3+1$ dimensions. In addition we have the isometries of ${\bf S}^5$ which
form $SU(4)\sim SO(6)$. This group is identical to the R-symmetry of
the ${\cal N}=4$ SYM theory. After including the fermionic generators
required by supersymmetry, the full isometry supergroup of the
$AdS_5\times {\bf S}^5$ background is $SU(2,2|4)$, which is identical to
the ${\cal N}=4$ superconformal symmetry.

The fact that after the compactification on $Y_5$ 
the string theory is 5-dimensional supports earlier ideas on the necessity
of the 5-th dimension to describe 4-d gauge theories\cite{Polyakov}.
The $z$-direction is dual to the energy scale of the gauge theory: 
small $z$ corresponds to the UV domain of the gauge theory, 
while large $z$ to the IR. 

In the AdS/CFT duality, type IIB strings
are dual to the chromo-electric flux lines in the gauge theory, 
providing a string theoretic set-up for calculating 
the quark anti-quark potential\cite{Malda}.
The quark and anti-quark are placed near the boundary of Anti-de Sitter space 
($z=0$), and the fundamental string connecting them is required
to obey the equations of motion following from the Nambu action.
The string bends into the interior ($z>0$), and the maximum value
of the $z$-coordinate increases with the separation $r$ between quarks.
An explicit calculation of the string action gives an 
attractive $q\bar q$ potential\cite{Malda}: 
\be
V(r) = -\frac{4\pi^2 \sqrt{\lambda }}{\Gamma \left(\frac{1}{4} \right)^4 r} \ .
\ee
Its Coulombic $1/r$ dependence is required by the conformal invariance of 
the theory. Historically, a dual string description was hoped for mainly in
the cases of confining gauge theories, where long confining
flux tubes have string-like properties. In a pleasant surprise, we now see that
a string description  applies to non-confining theories too,
due to the presence of extra dimensions with a warped metric.

\section{Semiclassical Spinning Strings vs. Highly Charged Operators}

A few years ago it was noted that the
AdS/CFT duality becomes particularly powerful when
applied to operators with large quantum numbers. One class of
such single-trace ``long operators'' are the BMN operators\cite{Berenstein}
that carry a large R-charge in the SYM theory and contain
a finite number of impurity insertions. The R-charge is dual
to a string angular momentum on the compact space $Y_5$.
So, in the BMN limit the relevant states are short closed strings
with a large angular momentum, and a small amount of vibrational 
excitation \cite{Gubser:2002tv}.
Furthermore, by increasing the number of impurities the string can be
turned into a large semi-classical object moving in $AdS_5\times Y_5$.
Comparing such objects with their dual long operators has become a very
fruitful area of research\cite{Tseytlinrev}. Work in this direction
has also produced a great deal of evidence that the ${\cal N}=4$
SYM theory is exactly integrable (see \cite{Beisert} for reviews).

Another familiar example of an operator with a large quantum number is the
twist-2 operator  carrying a large spin $S$,
\be              
{\rm Tr}\ F_{+ \mu} D_+^{S-2} F_+^{\ \ \mu} \ .
\ee
In QCD, such operators
play an important role in studies of deep inelastic scattering\cite{GrossWilczek}.
In general, the anomalous dimension of a
twist-2 operator grows logarithmically \cite{Korchemsky} for large spin $S$: 
\be \label{Universal}
\Delta-S = f(g) \ln S\ + O(S^0)\ ,
\ee
where $g ={\sqrt{g_{YM}^2 N}/
4\pi}$.
This was demonstrated early on at 1-loop 
order\cite{GrossWilczek}
and at two loops\cite{Floratos}, where a cancellation of $\ln^3 S$
terms occurs.  There are solid arguments that (\ref{Universal})
holds to all orders in perturbation 
theory\cite{Korchemsky,Sterman}.
The function of coupling $f(g)$ also
measures the anomalous dimension of a cusp in a light-like Wilson
loop \cite{Korchemsky}, and is of definite physical interest in QCD.

There has been significant interest in determining $f(g)$ in the
${\cal N}=4$ SYM theory, in which case we can consider operators in the $SL(2)$ sector, of the form
\be              
{\rm Tr}\ D^{S} Z^J + \ldots \ ,
\ee
where $Z$ is one of the complex scalar fields, the R-charge $J$ is the twist, 
and the dots serve as a reminder that the operator is a linear combination of 
such terms with the covariant derivatives acting on the scalars in all possible ways. 
The object dual to such a high-spin twist-2 operator
is a folded string\cite{Gubser:2002tv} spinning around the center of $AdS_5$; its generalization to large $J$ 
was found in \cite{FT}. The result (\ref{Universal}) is generally applicable when $J$ is held fixed while
$S$ is sent to infinity \cite{Belitsky:2006en}.
While the function $f(g)$ for ${\cal N}=4$ SYM is not the same as the cusp anomalous dimension for QCD, 
its perturbative expansion is in fact related to that in QCD
by the conjectured ``transcendentality principle'' \cite{Kotikov}.

The perturbative expansion
of $f(g)$ at small $g$ can be obtained from gauge theory,
but calculations in the
planar ${\cal N}=4$ SYM theory are quite formidable, and until
recently were available only up to 3-loop 
order\cite{Kotikov,Bern}:
\begin{equation}
f(g) = 8 g^2 -{8\over 3} \pi^2 g^4 + {88\over 45} \pi^4 g^6 +
O(g^8)\ .
\end{equation}
This ${\cal N}=4$ answer can be 
extracted \cite{Kotikov} from the corresponding 
QCD calculation \cite{Moch} using the transcendentality principle,
which states that each expansion coefficient has terms of definite
degree of transcendentality (namely the exponent of $g$ minus two), 
and that the QCD result contains the same terms (in addition to others which have lower degree of transcendentality).

The
AdS/CFT correspondence relates the large $g$
behavior of $f(g)$ to the energy of a folded string spinning around
the center of a weakly curved $AdS_5$ space\cite{Gubser:2002tv}. This gives
the prediction that $f(g)\to 4 g$ at strong coupling. The
same result was obtained from studying the cusp anomaly using
string theory methods\cite{Kruczenski}. Furthermore, the
semi-classical expansion for the spinning string energy predicts
the following correction\cite{FT}:
\begin{equation} \label{AdSpredict}
f(g) = 4 g - {3\ln 2\over \pi} + O(1/g)\ .
\end{equation}

An interesting problem is to smoothly match
these explicit predictions of
string theory for large $g$ to those of gauge theory at small $g$. During the past few years
methods of integrability in AdS/CFT\cite{Minahan}
(for reviews and more complete references, see \cite{Beisert}) have led to major progress in
addressing this question. In an impressive series of papers \cite{Eden,BHL,BES} an integral
equation that determines
$f(g)$ was proposed. This equation was obtained from the asymptotic Bethe ansatz for the 
$SL(2)$ sector by considering the limit $S \to \infty$ with $J$ finite, and extracting the 
piece proportional to $\ln S$, which is manifestly independent of $J$. 
Taking the spin to infinity, the discrete Bethe equations can be rewritten as an 
integral equation for the density of Bethe roots in rapidity space 
(though the resulting equation is most concisely expressed in terms of the variable $t$ that 
arises from performing a Fourier-transform).

The cusp anomalous dimension $f(g)$ is related to value of the 
fluctuation density $\hat\sigma(t)$ at $t=0$ \cite{Eden,BES,Belitsky:2006wg}: 
\be
f(g)=16g^2\hat\sigma(0)\ ,
\ee
where $\hat \sigma(t)$ is determined by the integral equation  
\be\label{IntEq}
\hat\sigma(t)=\frac{t}{e^t-1}\left(K(2gt,0)-4g^2\int_0^\infty dt'
K(2gt,2gt')\hat\sigma(t')\right).
\ee
The kernel $K(t,t') = K^{(m)}(t,t') + 
K^{(d)}(t,t')$
is the sum of the so-called main scattering kernel $K^{(m)}(t,t') = K_0(t,t') + K_1(t,t')$ and the "dressing kernel"
\be
K^{(d)}(t,t') = 8g^2 \int_0^\infty dt'' K_1(t,2gt'')
\frac{t''}{e^{t''}-1} K_0(2gt'',t') \ .
\ee
Here $K_0$ and $K_1$ can be expressed as the following sums of Bessel functions, which are even
and odd, respectively, under change of sign of both $t$ and $t'$:
\bea
K_0(t,t') &=&  {2\over t\, t'}\sum_{n=1}^{\infty} (2n-1) J_{2n-1}(t)J_{2n-1}(t')\ , \\
K_1(t,t') &=&  {2\over t\, t'}\sum_{n=1}^{\infty} (2n)
J_{2n}(t)J_{2n}(t')\ .
\eea
Including the dressing kernel \cite{BES} is of crucial importance
since it takes into account the
dressing phase in the asymptotic two-particle world-sheet S-matrix, which is the only function of rapidities appearing in the S-matrix that is not fixed a priori by the $PSU(2,2|4)$ symmetry of ${\cal N}=4$ SYM theory, and whose general form was
deduced in \cite{AFS,BK}. The perturbative expansion
of the phase starts at the 4-loop order, and at
strong coupling coincides with the earlier results from string
theory\cite{BHL,AFS,HL,Freyhult:2006vr,HM}.
The important requirement of crossing
symmetry\cite{Janik} is satisfied by this phase, and it also
obeys the transcendentality principle of \cite{Kotikov}. Thus, there is strong evidence that this
phase describes the exact magnon S-matrix at any
coupling, which constitutes important progress in the
understanding of the ${\cal N}=4$ SYM theory, and the AdS/CFT
correspondence.

An immediate check of the validity of the integral equation, which gives the expansion
\begin{equation} \label{planarexpand}
f(g) = 8 g^2 -{8\over 3} \pi^2 g^4 + {88\over 45} \pi^4 g^6 
-16 \left ({73\over 630} \pi^6 + 4\zeta (3)^2 \right ) g^8 + \, \mathcal{O}(g^{10}) \ ,\
\end{equation}
 was provided by the gauge theory calculation of the 4-loop, $\mathcal{O}(g^{8})$ term. 
In a remarkable
paper \cite{BCDKS}, which independently arrived at the same conjecture for the all-order expansion of
$f(g)$ as \cite{BES}, the 4-loop coefficient was calculated numerically by
on-shell unitarity methods in SYM theory, 
and was used to conjecture the analytic result of (\ref{planarexpand}).
Subsequently, the numerical precision was improved to produce agreement with
the analytic value within $0.001$ percent
accuracy \cite{Cachazo:2006az}.

In fact the perturbative
expansion of $f(g)$ can be obtained from the integral equation (\ref{IntEq})
to arbitrary order.
Although the expansion has a finite radius of convergence, as is
customary in planar theories, it is expected to determine the function completely.
But solving (\ref{IntEq}) for $\hat\sigma(t)$ at values of the coupling constant $g$ beyond the 
perturbative regime is not an easy task.
By expanding the fluctuation density as a Neumann series of Bessel functions, 
\be
\label{s-expa}
\hat\sigma(t) =\frac{t}{e^t-1} \sum_{n\geq 1}s_n\frac{J_n(2gt)}{2gt}\ ,
\ee
one can reduce the problem to an (infinite-dimensional) matrix problem, 
which can be consistently truncated and is thus amenable to numerical solution \cite{BBKS}.
This was shown to give $f(g)$ with high accuracy up
to rather large values of $g$, and the function was found to be monotonically increasing, smooth, and 
in excellent agreement with the linear asymptotics predicted
by string theory (\ref{AdSpredict}). The cross-over from perturbative to the linear 
behavior takes place at around $g\sim 1/2$, which is comparable to the radius of convergence $|g|=1/4$. 

Subsequently, the 
leading term in the strong coupling asymptotic expansion of the fluctuation density $\hat\sigma(t)$ was derived 
analytically\cite{AABEK,Kostov} from the integral equation. More recently, 
the complete asymptotic expansion of $f(g)$ was determined in an impressive paper 
\cite{Basso} (for further work, see \cite{Kost}). 
This expansion obeys its own transcendentality principle;
in particular, the coefficient of the $1/g$ term in (\ref{AdSpredict}) is
\be
-{{\rm K} \over  4\pi^2}\approx -0.0232\ ,
\ee
in agreement with the numerical work of
\cite{BBKS} (${\rm K}$ is the Catalan constant).
As a further check, this coefficient was reproduced analytically from a two-loop calculation
in string sigma-model perturbation theory \cite{Roiban}.

Thus, the cusp anomaly $f(g)$ is an
example of a non-trivial interpolation function for an observable not protected by supersymmetry
 that smoothly connects weak and strong coupling regimes, and tests the AdS/CFT 
correspondence at a very deep level.
The final form of $f(g)$ was arrived at using inputs from string theory, perturbative gauge
theory, and the conjectured exact integrability of planar ${\cal N}=4$ SYM.
Further tests of this proposal may be carried out perturbatively: in fact, the
planar expansion (\ref{planarexpand}) makes an infinite number of explicit analytic
predictions for higher loop coefficients. This shows how implications of string theory
may sometimes be tested via perturbative gauge theory.

\section{Thermal Gauge Theory from Near-Extremal D3-Branes}

\subsection*{Entropy}

An important black hole observable is the Bekenstein-Hawking (BH)
entropy, which is proportional to the area of the event horizon,
$S_{BH}= A_h/(4 G)$.
For the $3$-brane solution (\ref{metric}), 
the horizon is located at
$r=r_0$.
For $r_0>0$ the $3$-brane carries some excess
energy $E$ above its extremal value, and the
BH entropy is also non-vanishing. 
The Hawking temperature is then defined by
$ T^{-1} = \partial S_{BH}/\partial E$.

Setting $r_0\ll L$ in (\ref{metric}), we obtain
a near-extremal 3-brane geometry, whose 
Hawking temperature is found to be $T= r_0/(\pi L^2)$.
The small $r$ limit of this geometry is ${\bf S}^5$ times
a certain black hole in $AdS_5$.
The 8-dimensional ``area'' of the event horizon is
$ A_h = \pi^6 L^8 T^3 V_3 $,
where $V_3$ is the spatial volume of the D3-brane
(i.e. the volume of the $x^1, x^2, x^3$ coordinates).
Therefore, the BH entropy is\cite{gkp}
\begin{equation}
\label{bhe}
S_{BH}= {\pi^2\over 2} N^2 V_3 T^3
\ .
\end{equation}
This gravitational entropy of a near-extremal 3-brane
of Hawking temperature $T$ is to be identified
with the entropy
of ${\mathcal N}=4$ supersymmetric $U(N)$ gauge theory
(which lives on $N$ coincident D3-branes) heated up to
the same temperature.

The entropy of a free $U(N)$ ${\cal N}=4$
supermultiplet, which consists of the gauge field, $6 N^2$ massless
scalars and $4 N^2$ Weyl fermions, can be calculated using
the standard statistical mechanics of a massless gas (the black body 
problem), and the answer is
\be \label{ffc}
S_0= {2 \pi^2\over 3} N^2 V_3 T^3
\ .
\ee
It is remarkable that the 3-brane geometry
captures the $T^3$ scaling characteristic of a conformal
field theory (in a CFT this scaling is guaranteed by the extensivity of
the entropy and the absence of dimensionful parameters).
Also, the $N^2$ scaling indicates the presence of $O(N^2)$
unconfined degrees of freedom, which is exactly what we expect in
the ${\cal N}=4$ supersymmetric $U(N)$ gauge theory.
But what is the explanation of the relative factor of $3/4$ between $S_{BH}$ and
$S_0$?
In fact, this factor
is not a contradiction but rather a {\it prediction} about the strongly
coupled ${\cal N}=4$ SYM theory at finite temperature. 
As we argued above, the supergravity calculation of the
BH entropy, (\ref{bhe}),
is relevant to the $\lambda\rightarrow\infty$ limit of the 
${\cal N}=4$ $SU(N)$ gauge theory,
while the free field calculation, (\ref{ffc}), 
applies to the $\lambda\rightarrow
0$ limit. Thus, the relative factor of $3/4$ is not a discrepancy:
it relates two different limits of the theory. 
Indeed, on general field theoretic
grounds, in the `t Hooft large $N$
limit the entropy is given by\cite{GKT}
\be
S= {2 \pi^2\over 3} N^2 f_e(\lambda) V_3 T^3
\ .\ee
The function $f_e$ is certainly not constant: Feynman graph calculations
valid for small $\lambda=g_{\rm YM}^2 N$ give\cite{Foto}
\be \label{weak} 
f_e(\lambda) = 1 - {3\over 2\pi^2} \lambda 
+{3+\sqrt 2\over \pi^3} \lambda^{3/2} + \ldots
\ee
The BH entropy in supergravity, (\ref{bhe}),
is translated into the prediction that 
\be
\lim_{\lambda\rightarrow \infty}
f_e(\lambda ) = {3\over 4}\ .
\ee 
A string theoretic calculation of the leading correction
at large $\lambda$ gives\cite{GKT}
\be \label{strong} 
f_e(\lambda) = {3\over 4}  +  {45\over 32}\zeta(3)
\lambda^{-3/2} + \ldots
\ee
These results are consistent with a monotonic function
$f_e(\lambda)$ which decreases from 1 to 3/4 as $\lambda$ is increased
from 0 to $\infty$. The 1/4 deficit compared to the free field value
is a strong coupling effect predicted by the AdS/CFT correspondence.

It is interesting that similar deficits have been observed in 
lattice simulations of deconfined non-supersymmetric 
gauge theories\cite{Karsch,Gavai,Bringoltz}.
The ratio of entropy to its free field value,
calculated as a function of the temperature, is found to
level off at values around
$0.8$ for $T$ beyond 3 times the deconfinement temperature $T_c$.
This is often interpreted as the effect of a sizable coupling.
Indeed, for $T= 3 T_c$, the lattice estimates \cite{Gavai}  indicate
that $g_{\rm YM}^2 N \approx 7$.
This challenges
an old prejudice that the quark-gluon plasma is inherently weakly coupled.
We now turn to
calculations of the shear viscosity where strong coupling effects
are more pronounced. 

\subsection*{Shear Viscosity}

The shear viscosity $\eta$ may be read off from the form of
the stress-energy tensor 
in the local rest frame of the fluid where $T_{0i}=0$:
\be
T_{ij} = p \delta_{ij} -
\eta (\partial_i u_j + \partial_j u_i -{2\over 3} \delta_{ij}
\partial_k u_k )
\ ,
\ee
where $u_i$ is the 3-velocity field. 
The viscosity can be also determined\cite{Policastro}
through the Kubo formula
\be
\eta = \lim_{\omega\to 0} {1\over 2\omega}
\int dt d^3 x e^{i\omega t} \langle [T_{xy}(t,\vec x), T_{xy}(0,0)] \rangle
\ee
For the ${\cal N}=4$ supersymmetric YM theory this 
2-point function may be computed from absorption of a low-energy
graviton $h_{xy}$
by the 3-brane metric\cite{absorption}.
Using this method, it was found\cite{Policastro}
that at very strong coupling 
\be
\eta= {\pi\over 8} N^2 T^3
\ ,
\ee
which implies 
\be \label{bound}
{\eta\over s} = {\hbar \over 4\pi}
\ee
after 
$\hbar$ is restored in the calculation (here 
$s = S/V_3$ is the entropy density). This is much smaller than what 
perturbative estimates imply.
Indeed, at weak coupling $\eta/s$ is very
large, $\sim {\hbar \lambda^{-2} / \ln (1/\lambda)}$ \cite{Huot},
and there is evidence that it 
decreases monotonically as the coupling is increased\cite{Buchel}.

The saturation of $\eta/s$ at some value of order $\hbar$ 
is reasonable on general physical grounds\cite{Kovtun}. 
The shear viscosity $\eta$ is of order the 
energy density times quasi-particle mean free time $\tau$. 
So $\eta/s $ is of order
of the energy of a quasi-particle times its mean free time, 
which is bounded from below by the uncertainty principle
to be some constant times $\hbar$.
These considerations prompted a suggestion \cite{Kovtun}
that $\hbar/(4\pi)$ is
the lower bound on $\eta/s$. However, the generality of this bound was called into question in \cite{Cohen}.
Recently it was shown \cite{Kats} that for special
large $N$ gauge theories whose AdS duals contain
D7-branes, the leading $1/N$ correction to (\ref{bound}) is negative; thus, the bound can be violated
even for theories that have a holographic description. 

Nevertheless, (\ref{bound}) applies to the strong coupling limit of a large class of
gauge theories that have gravity
duals. Its important physical implication is that a low value of
$\eta/s$ is a consequence of strong coupling.
For known fluids (e.g. helium, nitrogen, water) $\eta/s$
 is considerably higher than (\ref{bound}).
On the other hand,
the quark-gluon plasma produced at RHIC is believed  
to have a very low $\eta/s$  
\cite{Teaney,Hirano}, with some recent estimates \cite{Romatschke} suggesting that
it is below (\ref{bound}). 
Lattice studies of pure glue gauge theory \cite{Meyer}) also lead to low values of $\eta/s$.
This suggests that, at least for $T$ near $T_c$, the theory is sufficiently strongly
coupled.
Indeed, a new term, sQGP, which stands for 
``strongly coupled quark-gluon plasma,''
has been coined to describe the deconfined state observed at 
RHIC\cite{Gyulassy,Shuryak} (a somewhat different term,
``non-perturbative quark-gluon plasma'', was advocated in \cite{Pisarsky}).
As we have reviewed, the gauge-string duality is a 
theoretical laboratory which allows
one to study some extreme examples
of such a new state of matter. These include the thermal
${\cal N}=4$ SYM theory at very strong `t~Hooft coupling, as well as other gauge theories
which have less supersymmetry and exhibit confinement at low temperature. 
An example of such a gauge theory, whose dual is the warped deformed conifold, will
be discussed in section \ref{deformed}.

Lattice calculations indicate that the deconfinement temperature
$T_c$ is around $175$ MeV, and  
the energy density is $\approx 0.7$ GeV/fm$^3$,
around 6 times the nuclear energy density.
RHIC has reached energy densities around $14$ GeV/fm$^3$, 
corresponding to $T \approx 2 T_c$.
Furthermore, heavy ion collisions at the
LHC are expected to reach temperatures up to $5 T_c$.  
Thus, RHIC and LHC should provide a great deal of new information about the sQGP,
which can be compared with calculations based on gauge-string duality.
For a more detailed discussion of the recent theoretical work in this direction, see
U. Wiedemann's lectures in this volume.

\section{Warped Conifold and its Resolution}
\label{resolved}

To formulate the AdS/CFT duality at zero temperature, but with a reduced amount of supersymmetry,
we may place the stack of D3-branes at the tip of
a 6-dimensional Ricci flat cone,\cite{ks,Kehag,KW}
whose base is a 5-dimensional compact Einstein space $Y_5$.
The metric of such a cone is $dr^2 + r^2 ds_Y^2$; therefore,
the 10-d metric produced by the D3-branes is obtained from (\ref{geom})
by replacing $d\Omega_5^2$, the metric on ${\bf S}^5$,
by $ds_Y^2$, the metric on $Y_5$. In the $r\to 0$ limit we then find
the space $AdS_5\times Y_5$ as the candidate dual of the CFT
on the D3-branes placed at the tip of the cone.
The isometry group of $Y_5$ is smaller than $SO(6)$,
but $AdS_5$ is the ``universal'' factor present in the dual
description of any large $N$ CFT, making the $SO(2,4)$ conformal
symmetry geometric.

To obtain gauge theories with ${\cal N}=1$ superconformal symmetry the
Ricci flat cone must be a Calabi-Yau 3-fold
\cite{KW,Morrison} whose base
$Y_5$ is called a
Sasaki-Einstein space.
Among the simplest examples of these is $Y_5 = T^{1,1}$. The corresponding
Calabi-Yau cone is called the conifold. Much of the work on gauge-string dualities
originating from D-branes on the conifold is reviewed in the 2001 Les Houches lectures
\cite{Herzog}. Here we will emphasize the progress that has taken place since then, in particular
on issues related to breaking of the $U(1)$ baryon number symmetry. 

 The conifold is a singular non-compact Calabi-Yau three-fold \cite{Candelas}. Its importance
arises from the fact that the generic singularity in a Calabi-Yau
three-fold locally looks like the conifold, 
described by the quadratic equation in ${\bf C}^4$:
\be\label{conifold}
 z_1^2 + z_2^2 + z_3^2 + z_4^2 = 0\ .
\ee
 This homogeneous equation defines a real cone over a 5-dimensional manifold. For the cone to be
 Ricci-flat
the 5d base must be an Einstein manifold ($R_{ab} = 4 g_{ab}$).
 For the conifold \cite{Candelas}, the topology of the base can be shown to be 
${\bf S}^2 \times {\bf S}^3$ and it is
 called $T^{1,1}$, with the following Einstein metric \cite{Romans}:
\begin{eqnarray} \label{T11metric}
  d\Omega_{T^{1,1}}^2 &=&  \frac{1}{9}  \left( d\psi + \cos \theta_1 d\phi_1 + \cos \theta_2 d\phi_2 \right)^2  \nonumber \\
 & & + \frac{1}{6}  (d\theta_1^2 + \sin^2 \theta_1 d\phi_1^2) + \frac{1}{6} (d\theta_2^2 + \sin^2 \theta_2 d\phi_2^2)\ .
\end{eqnarray}
$T^{1,1}$ is a homogeneous space, being the coset
$SU(2)\times SU(2)/U(1)$. 
The metric on the cone is then $ds^2 = dr^2 + r^2
d\Omega_{T^{1,1}}^2$. 

We may introduce two other types  of complex coordinates on the
conifold, $w_a$ and $a_i,b_j$, as follows:
\begin{eqnarray} \label{coords}
 Z &=& \left( \begin{array}{cc}  z_3 + i z_4 & z_1 - i z_2 \\ z_1 +i z_2 & -z_3 +i z_4 \\ \end{array} \right) \nonumber
 = \left(\begin{array}{cc}w_1 & w_3 \\ w_4& w_2 \end{array} \right)
= \left( \begin{array}{cc}  a_1 b_1 & a_1 b_2 \\ a_2 b_1 & a_2 b_2 \\ \end{array} \right) \\
&=& r^\frac{3}{2} \left(\begin{array}{cc} - c_1 s_2 \; e^{\frac{i}{2} ( \psi + \phi_1-\phi_2)} &
    c_1 c_2 \; e^{\frac{i}{2} ( \psi+ \phi_1 +\phi_2)} \\
                  - s_1 s_2 \; e^{\frac{i}{2} ( \psi - \phi_1 -\phi_2)} & s_1 c_2 \;e^{\frac{i}{2}
                  ( \psi - \phi_1+\phi_2)} \end{array} \right)\ ,
\end{eqnarray}
where $c_i = \cos \frac{\theta_i}{2},\ s_i = \sin
\frac{\theta_i}{2}$ (see \cite{Candelas} for other details on the
$w,z$ and angular coordinates). The equation defining the conifold
is now $\det Z = 0$.

The $a,b$ coordinates above will be of particular interest to us because the symmetries of the conifold are most apparent in
this basis. The conifold equation has $SU(2)\times SU(2) \times
U(1)$ symmetry since under these symmetry transformations, \be
\det L Z R^{\dagger} = \det e^{i \alpha} Z = 0. \ee 
This is also a
symmetry of the metric presented above where each $SU(2)$ acts on
$\theta_i,\phi_i,\psi$ (thought of as Euler angles on ${\bf S}^3$) while
the $U(1)$ acts by shifting $\psi$. This symmetry can be
identified with $U(1)_R$, the R-symmetry of the dual gauge theory,
 in the conformal case.
The action of the $SU(2) \times SU(2) \times U(1)_R$ symmetry on
$a_i,b_j$ defined in (\ref{coords}) is given by: 
\bea SU(2) \times SU(2)
\mbox{  symmetry}&:& \quad
 \left( a_1 \atop a_2 \right)  \rightarrow L \; \left( a_1 \atop a_2 \right) , \\
&& \quad\, \left( b_1 \atop b_2 \right)  \rightarrow R \;
\left( b_1 \atop b_2 \right) , \\
 \mbox{R-symmetry}&:& \quad (a_i,b_j) \rightarrow e^{i \frac{\alpha}{2}} (a_i,b_j)
\ ,\eea 
i.e.~$a$ and $b$ transform as $(1/2,0)$ and $(0,1/2)$
under $SU(2) \times SU(2)$ with R-charge $1/2$ each.
We can thus describe the singular conifold as the manifold parametrized by
$a,b$, but from (\ref{coords}), we see that there is some redundancy
in the $a,b$ coordinates. Namely, the transformation \be
\label{removeredun}
  a_i  \rightarrow \; \lambda \; a_i  \ ,\quad  b_j  \rightarrow \; \frac{1}{\lambda} \; b_j \ , \quad(\lambda \in \mathbb{C})\ ,
\ee
 give the same $z,w$ in (\ref{coords}).
 We impose the constraint $|a_1|^2 + |a_2|^2 - |b_1|^2 -|b_2|^2 = 0$ to fix the magnitude in the
 above transformation. To account for the remaining phase, we describe the singular conifold as
 the quotient of the $a,b$ space with the above constraint by the relation
 $ a \sim e^{i\alpha} a, b \sim e^{-i\alpha} b$.

The importance of the coordinates $a_i, b_j$ is that in the gauge theory on D3-branes
at the tip of the conifold they are promoted to chiral superfields.
The low-energy
gauge theory on $N$ D3-branes is a ${\cal
N}=1$ supersymmetric $SU(N)\times SU(N)$ gauge theory with
bi-fundamental chiral superfields $A_i$, $B_j$ $(i,j=1,2)$ in the
$ ({\bf N},\overline{\bf{N}})$ and
$ (\overline{{\bf N}},\bf{N})$
 representations of the gauge groups,
respectively \cite{KW,Morrison}. The superpotential for this gauge
theory is 
\be \label{suppot}
W \sim {\rm Tr} \det A_i B_j = {\rm Tr}\ (A_1 B_1 A_2 B_2 -
A_1 B_2 A_2 B_1)
\ .
\ee 
The continuous global symmetries of this theory
are $SU(2) \times SU(2) \times U(1)_B \times U(1)_R$, where the $SU(2)$
factors act on
$A_i$ and $B_j$ respectively, $U(1)_B$ is a baryonic
symmetry under which the $A_i$ and $B_j$ have opposite charges, and $U(1)_R$ is the R-symmetry with charges of the same sign $R_A=R_B=\frac{1}{2}$.
This assignment ensures that $W$ is marginal, and one can also show
that the gauge couplings do not run. Hence this theory is
superconformal for all values of gauge couplings and superpotential
coupling \cite{KW,Morrison}.

A simple way to understand the resolution of the conifold is to deform the modulus constraint above into
\begin{equation}
\label{resolution}
|b_1|^2+|b_2|^2-|a_1|^2-|a_2|^2=u^2\ ,
\end{equation}
where $u$ is a real parameter which controls the resolution. The resolution corresponds 
to a blow up of the ${\bf S}^2$ at the bottom of the conifold. 
In the dual gauge theory turning on $u$ corresponds to a particular choice of vacuum \cite{Klebanov:1999tb}. 
After promoting the $a,b$ fields into the bi-fundamental chiral superfields
of the dual gauge theory, we can define the operator $\mathcal{U}$ as
\begin{equation}
\mathcal{U}=\frac{1}{N}{\rm{Tr}} (B_1^{\dagger}B_1+B_2^{\dagger}B_2-A_1^{\dagger}A_1-A_2^{\dagger}A_2)\ .
\end{equation}
Thus, the warped singular conifolds correspond to gauge theory vacua where 
$\langle \mathcal{U}\rangle=0$, while the warped
resolved conifolds correspond to vacua where $\langle \mathcal{U}\rangle\ne 0$. 
In the latter case, some VEVs for the bi-fundamental fields 
$A_i,B_j$ must be present. Since these 
fields are charged under the $U(1)_B$ symmetry, 
the warped resolved conifolds  
correspond to vacua where this symmetry is broken \cite{Klebanov:1999tb}.

A particularly simple choice is to give a diagonal VEV to only one of the 
scalar fields, say, $B_2$. 
As seen in  \cite{Klebanov:2007us}, this choice breaks the $SU(2)\times SU(2)\times U(1)_B\times U(1)_R$ 
symmetry of the CFT down to $SU(2)\times U(1)\times U(1)$. 
The string dual is given by a warped resolved
conifold 
\begin{equation}
ds^2=h^{-1/2} \eta_{\mu\nu} dx^\mu dx^\nu +h^{1/2}ds_6^2\ .
\end{equation}
The explicit form of the Calabi-Yau metric of the resolved conifold is given by \cite{Pando:2000sq}
\begin{eqnarray}
ds_6^2&=&K^{-1}dr^2+\frac{1}{9}Kr^2\Big(d\psi+\cos\theta_1 d\phi_1+\cos\theta_2 d\phi_2\Big)^2 \nonumber \\
&&
+\frac{1}{6}r^2(d\theta_1^2+\sin^2\theta_1d\phi_1^2)
+\frac{1}{6}(r^2+6u^2)(d\theta_1^2+\sin^2\theta_2d\phi_2^2)\ , 
\end{eqnarray}
where
\begin{equation}
K=\frac{r^2+9u^2}{r^2+6u^2}\ .
\end{equation}
The $N$ $D3$-branes sourcing the warp factor are
located at the north pole of the finite ${\bf S}^2$, i.e. at $r=0, \theta_2=0$. 
The warp factor is the Green function on the resolved conifold with this source
\cite{Klebanov:2007us}:
\begin{equation} \label{KMwarp}
h(r, \theta_2)=
L^4\sum_{l=0}^{\infty}(2l+1)H_l(r)P_l(\cos\theta_2)\ .
\end{equation}
Here $L^4={27\pi\over 4} g_s N(\alpha')^2$, 
$P_l(\cos\theta)$ is the $l$-th Legendre polynomial, and
\begin{equation}
H_l=\frac{2C_{\beta}}{9 u^2 r^{2+2\beta}}\quad_2F_1\Big(\beta,1+\beta,1+2\beta;-\frac{9u^2}{r^2}\Big)\ ,
\end{equation}
with the coefficients $C_{\beta}$ and $\beta$ given by
\begin{equation}C_{\beta}=\frac{(3u)^{2\beta}\Gamma(1+\beta)^2}{\Gamma(1+2\beta)}\ ,\qquad \beta=\sqrt{1+\frac{3}{2}l(l+1)}\ .
\end{equation}

Far in the IR the gauge theory flows to the $\mathcal{N}=4$ $SU(N)$ SYM theory, 
as evidenced by the appearance
of an $AdS_5\times {\bf S}^5$ throat near the location of the stack of the $D3$-branes. 
This may be verified in the gauge theory as follows.
The condensate $B_2 = {u}\, 1_{N\times N}$ breaks the
$SU(N) \times SU(N)$ gauge group down to $SU(N)$, all the chiral fields now transforming in the
adjoint of this diagonal group. 
After substituting this classical value for $B_2$,
the quartic superpotential (\ref{suppot}) reduces to the cubic $ {\cal N}=4$ form,
\be
  W \sim {\rm Tr} (A_1[B_1,A_2])\ .
\ee
This confirms that the gauge theory flows to the $ {\cal N}=4$ $SU(N)$ SYM theory.
The gauge theory also contains an interesting additional sector coupled to 
this infrared CFT; in particular, it contains global strings due to the breaking
of the $U(1)_B$ symmetry \cite{KMRW}.

\subsection*{Baryonic Condensates and Euclidean D3-branes}

The gauge invariant order parameter for the breaking of 
$U(1)_B$ is ${\rm det} B_2$.
Let us review the calculation of this baryonic VEV using the dual string theory on
the warped resolved conifold background \cite{Klebanov:2007us}.

The objects in $AdS_5\times T^{1,1}$ that are dual to baryonic operators
are D3-branes wrapping 3-cycles in $T^{1,1}$ \cite{GK}. Classically, the
3-cycles dual to the baryons made out of the $B$'s are located at fixed
$\theta_2$ and $\phi_2$, while quantum mechanically one has to carry out
collective coordinate quantization and finds wave functions of spin $N/2$
on the 2-sphere. 

To calculate VEVs of such baryonic operators we need to consider the action of a Euclidean D3-brane 
whose world volume ends at large $r$ on the 3-sphere at fixed $\theta_2$ and $\phi_2$. 
The D3-brane action should be integrated up to a radial cut-off $r$, and we identify 
$e^{-S(r)}$ with the field $\varphi(r)$ dual to the baryonic operator.
Close to the boundary, a field $\varphi$ dual to an operator of dimension
$\Delta$ in the AdS/CFT correspondence behaves as
\begin{equation} \label{sourceVEV}
\varphi(r) \rightarrow \varphi_0\, r^{\Delta-4} + A_\varphi\, r^{-\Delta}\ ,
\end{equation}
Here $A_\varphi$ is the operator expectation value \cite{KW2}, and
$\varphi_0$ is the source for it. There are no sources added for baryonic operators, hence
we will find that $\varphi_0=0$, but the term scaling as
$r^{-\Delta}$ is indeed present in
$e^{-S(r)}$.

The Born-Infeld action of the D3-brane is given by
\be
S_{BI}= T_3 \int d^4\xi \sqrt{ g} 
\ ,
\ee
where $g_{\mu\nu}$ is the metric induced on the D3 world-volume. 
The smooth 4-chain which solves the BI equations of motion subject to our boundary conditions 
is located at fixed
$\theta_2$ and $\phi_2$, and spans the $r$, $\theta_1$, $\phi_1$ and $\psi$ directions. 
Using the D3-brane tension 
\be T_3= {1\over g_s (2\pi)^3 (\alpha')^2 }
\ ,
\ee
we find 
\be
S_{BI}= {3N\over 4}\int_0^{r} d\tilde r \tilde r^3 h (\tilde r,\theta_2) 
\ .
\ee

Using the expansion (\ref{KMwarp}), we note that
the $l=0$ term needs to be evaluated separately since it contains
a logarithmic divergence:
\be
\int_0^{r} d\tilde r \tilde r^3 H_0 (r)= {1\over 4} + {1\over 2}
\ln \left (1+ {r^2\over 9 u^2} \right )
\ .
\ee
For the $l>0$ terms the integral converges and we find
the simple result \cite{Klebanov:2007us}
\be
\int_0^{\infty} d\tilde r \tilde r^3 \sum_{l=1}^\infty H_l (\tilde r) P_l (\cos \theta_2)
={2\over 3} (-1 -2 \ln [\sin (\theta_2/2)])\ .
\ee
This expression is recognized as the Green's function on a sphere.
Combining the results, and taking $r\gg u$, we find
\be
e^{-S_{BI}} = \left ( {3 e^{5/12} u\over r }\right )^{3N/4} 
\sin^N (\theta_2/2)\ .
\ee

In \cite{GK} it was argued that the wave functions
of $\theta_2, \phi_2$, which arise though
the collective coordinate quantization of
the D3-branes wrapped over the 3-cycle $(\psi,\theta_1,\phi_1)$, correspond to
eigenstates of a charged particle on ${\bf S}^2$ in the presence of a 
charge $N$ magnetic monopole. Taking the gauge potential 
$A_\phi =N (1+\cos \theta)/2, \ A_\theta=0$ we find that 
the ground state wave function $\sim \sin^N (\theta_2/2)$ carries the $J=N/2$ and $m=-N/2$ quantum numbers.\footnote{In
a different gauge this wave function would acquire a phase. In the string
calculation the phase comes from the purely imaginary Chern-Simons term in
the Euclidean D3-brane action.} 
These are the $SU(2)$ quantum numbers of ${\rm det} B_2$.
Therefore, the angular dependence of $e^{-S}$ identifies ${\rm det} B_2$
as the only operator that acquires a VEV, in agreement with the gauge theory. 

The power of $r$ indicates that the operator dimension is $\Delta=3N/4$, which
is indeed
the exact dimension of the baryonic operators \cite{GK}.
The VEV depends on the parameter $u$
as $\sim u^{3N/4}$. This is not the same as the classical scaling that would give
${\rm det} B_2\sim u^N $. The classical scaling is not obeyed because we are dealing with a
strongly interacting gauge theory where the baryonic operator acquires an anomalous dimension.

\subsection*{Goldstone Bosons and Global Strings}

Since the $U(1)_B$ symmetry is broken spontaneously, the spectrum of the
theory must include a
Goldstone boson. Its gravity dual is a normalizable RR 4-form fluctation
around the warped resolved conifold \cite{KMRW}:
\begin{equation}
\delta F^{(5)}=(1+\star)d (a_2(x)\wedge W)\ .
\end{equation}
Here $W$ is a closed 2-form inside the warped resolved conifold,
\begin{equation}
W=\sin\theta_2 d\theta_2\wedge d\phi_2+d(f_1g^5+f_2\sin\theta_2d\varphi_2)\ ,
\end{equation}
where $f_1, f_2$ are functions of $r,\theta_2$.
The equations of motion reduce to
\begin{equation}\label{eoma_2}
d\star_4da_2=0\ ,
\end{equation}
provided $W$ satisfies
\begin{equation}
d(h\star_6W)=0\ ,
\end{equation}
where $\star_4$, $\star_6$ are the Hodge duals with respect to the
unwarped Minkowski and resolved conifold metrics, respectively.
This gives coupled PDE's for $f_1$ and $f_2$. Their solution can be obtained
through minimizing a positive definite functional subject to certain boundary conditions
\cite{KMRW}.
Introducing the Goldstone boson field $p(x)$ through $\star_4 da_2=dp$,
we note that the fluctuation in the 5-form field strength reads
\begin{equation}
\label{RRpert}
\delta F^{(5)}=da_2\wedge W+dp\wedge h\star_6W\ .
\end{equation}
The corresponding fluctuation of the 4-form potential is
\begin{equation}
\label{C4}
\delta C^{(4)}=
a_2(x)\wedge W + p\wedge h\star_6W\ .
\end{equation}

In addition to the existence of a Goldstone boson, a hallmark of a broken $U(1)$ symmetry is the 
appearance of ``global'' strings. The Goldstone boson has a non-trivial monodromy around such a string,
thus
giving it a logarithmically divergent energy density.
On  
the string side of the duality these global strings are $D3$-branes wrapping the
2-sphere at the bottom of the warped resolved conifold \cite{KMRW}.
The Goldstone bosons and the global strings interact with
the $\mathcal{N}=4$ SYM theory that appears far in the infrared. 
The coupling of such an extra sector and an infrared CFT
is an interesting fact reminiscent of the unparticle physics scenarios \cite{Georgi:2007ek}.

\section{Deformation of the Conifold}
\label{deformed}

We have seen above that the singularity of the cone over $T^{1,1}$ can be replaced by
an $\bf{S}^2$ through 
resolving the conifold (\ref{conifold}) as in (\ref{resolution}). An alternative supersymmetric
blow-up,
which replaces the singularity by an $\bf{S}^3$,
is the deformed
conifold \cite{Candelas}
\begin{equation} \label{dconifold}
z_1^2 + z_2^2 + z_3^2 + z_4^2 =
\varepsilon^2\ .
\end{equation}
To achieve the deformation, one needs to turn on $M$ units of RR 3-form flux. This modifies
the dual gauge theory to 
${\cal N}=1$ supersymmetric
$SU(N) \times SU(N+M)$ theory coupled to chiral superfields
$A_1, A_2$
in the $ ({\bf N},\overline{\bf{N+M}})$ representation, and
$B_1, B_2$
in the $ (\overline{{\bf N}},\bf{N+M})$ representation.
Indeed, in type IIB string theory D5-branes source the 7-form field
strength from the Ramond-Ramond sector, which is Hodge dual to
the 3-form field strength. Therefore,
the $M$ wrapped D5-branes
create $M$ flux units of this field strength
 through the 3-cycle in the conifold;
this number
is dual to the difference between the numbers of colors
in the two gauge groups.
Thus, unlike the resolution, the deformation
cannot be achieved in the context of the $SU(N)\times SU(N)$ gauge theory.

The 10-d metric
takes the following form \cite{KS}:
\be \label{specans}
ds^2_{10} =   h^{-1/2}(\tau)\,  \eta_{\mu\nu} dx^\mu dx^\nu
 +  h^{1/2}(\tau)\, ds_6^2 \ ,
\ee
where $ds_6^2$ is the Calabi-Yau metric of the deformed conifold:
\bea \label{metricd}
ds_6^2 &=& {\varepsilon^{4/3} \over 2} K(\tau)
\Big[\sinh^2 \left({\tau\over 2}\right)  \left[(g^1)^2 + (g^2)^2\right] \\ \nonumber && \qquad\qquad + \cosh^2 \left({\tau\over 2}\right) \left[(g^3)^2 + (g^4)^2\right] + {1\over 3 K^3(\tau)} \left[d\tau^2 + (g^5)^2\right]\Big]
\ , 
\eea
where
\be
K(\tau)= { (\sinh\tau \cosh\tau - \tau)^{1/3}\over \sinh \tau}
\ .
\ee
For $\tau \gg 1$ we may introduce another radial coordinate $r$ defined by
\be \label{changeofc}
r^2 = {3\over 2^{5/3}} \varepsilon^{4/3} e^{2\tau/3}\ ,
\ee
and in terms of this coordinate we find
$ ds_6^2 \rightarrow dr^2 + r^2 ds^2_{T^{1,1}}$.

The basis one-forms $g^i$ in terms of which this metric is diagonal are defined by
\begin{eqnarray}
g^1 &\equiv& {e_2-\epsilon_2\over \sqrt{2}}\ , \quad g^2 \equiv {e_1-\epsilon_1\over \sqrt{2}}\ ,\\
g^3 &\equiv& {e_2+\epsilon_2\over \sqrt{2}}\ , \quad g^4 \equiv {e_1+\epsilon_1\over \sqrt{2}}\ ,\\
g^5 &\equiv& \epsilon_3+\cos\theta_1d\phi_1\ ,
\end{eqnarray}
where the $e_i$ are one-forms on ${\bf S}^2$
\begin{eqnarray}
e_1\equiv d\theta_1\ ,\qquad e_2\equiv  - \sin\theta_1 d\phi_1\ ,
\end{eqnarray}
and the $\epsilon_i$ a set of one-forms on ${\bf S}^3$
\begin{eqnarray}
\epsilon_1 &\equiv&\sin\psi\sin\theta_2 d\phi_2+\cos\psi d\theta_2\ , \\
\epsilon_2 &\equiv&   \cos\psi\sin\theta_2 d\phi_2 - \sin\psi d\theta_2\ ,\\
\epsilon_3 &\equiv& d\psi + \cos\theta_2 d\phi_2\ .
\end{eqnarray}
The NSNS two-form is given by
\bea \nonumber
B^{(2)} = \frac{g_s M\alpha'}{2}\frac{\tau \coth\tau - 1}{\sinh\tau}
\left[\sinh^2\left(\frac{\tau}{2}\right)\, g^1\wedge g^2 + \cosh^2\left(\frac{\tau}{2}\right)\, g^3\wedge g^4 \right]\\
\eea
and the RR fluxes are most compactly written as
\begin{eqnarray} \label{fluxcompact}
F^{(3)} &=& \frac{M\alpha'}{2}\left\{g^3\wedge g^4 \wedge g^5 +
d \left[\frac{\sinh\tau - \tau}{2 \sinh\tau}(g^1\wedge g^3+g^2\wedge g^4) \right] \right\}, \\ \label{fluxcompact2}
\tilde{F}^{(5)} &=& dC^{(4)} + B^{(2)} \wedge F^{(3)} = (1+\ast)\,(B^{(2)} \wedge F^{(3)})\ .
\end{eqnarray}
Note that the complex three-form field of this BPS supergravity solution is 
imaginary self dual: 
\be
\star_6 G_3 = i G_3\ , \qquad\qquad G_3 = F_3 - {i\over g_s} H_3\ ,
\ee
where $\star_6$ again denotes the Hodge dual with respect to the unwarped metric
$ds_6^2$.
This guarantees that the dilaton is constant, and
we set $\phi=0$.

The above expressions for the NSNS- and RR-forms follow by making a simple ansatz consistent 
with the symmetries of the problem, and solving a system of differential equations, 
which owing to the supersymmetry of the problem are only first order \cite{KS}. 
The warp factor is then found to be completely determined up to an additive constant, 
which is fixed by demanding that it go to zero at large $\tau$:
\begin{eqnarray} \label{deformedwarp}
h(\tau) &=& (g_s M \alpha')^2 2^{2/3} \varepsilon^{-8/3} I(\tau)\  ,\\
I(\tau) &\equiv& 2^{1/3}\int_{\tau}^\infty dx \frac{x \coth x - 1}{\sinh^2 x}(\sinh x \cosh x - x)^{1/3}\ .
\end{eqnarray}
For small $\tau$ the warp factor approaches a finite constant since $I(0)\approx 0.71805$.
This implies confinement because the chromo-electric flux tube, described by a fundamental
string at $\tau=0$, has tension
\be \label{conften}
T_s= {1\over 2\pi \alpha' \sqrt{h(0)} } \ .
\ee

The KS solution \cite{KS} is $SU(2)\times SU(2)$ symmetric and the expressions above 
can be written in an explicitly $SO(4)$ invariant way. 
It also possesses a $\mathbb{Z}_2$ symmetry ${\cal I}$, which exchanges $(\theta_1,\phi_1)$ with $(\theta_2,\phi_2)$
accompanied by the action of $-I$ of SL(2, $\mathbb{Z}$), changing the signs of the three-form fields.

Examining the metric $ds_6^2$ for $\tau=0$ we see that it degenerates into
\be
d\Omega_3^2= {1\over 2} \varepsilon^{4/3} (2/3)^{1/3}
\left[ {1\over 2} (g^5)^2 + (g^3)^2 + (g^4)^2 \right]
\ ,
\ee
which is the metric of a round $\bf{S}^3$, while
the $\bf{S}^2$ spanned by the other two angular coordinates, and fibered
over the $\bf{S}^3$, shrinks to zero size. In the ten-dimensional metric (\ref{specans}) this appears multiplied by a factor of $h^{1/2}(\tau)$, and thus the radius squared of the three-sphere at the tip of the conifold is of order $g_s M \alpha'$. Hence for $g_s M$ large, the curvature of the $\bf{S}^3$, and in fact everywhere in this manifold, is small and the supergravity
approximation reliable.

The field theoretic interpretation of the KS solution is unconventional.
After a finite distance along the RG flow, the
$SU(N+M)$ group undergoes a Seiberg
duality transformation\cite{Seiberg}. After this transformation,
and an interchange of the two gauge groups,
the new gauge theory is $SU(\tilde N )\times SU(\tilde N+ M)$
with the same bi-fundamental field content and superpotential, and with $\tilde N=N-M$.
The self-similar structure of the gauge theory under the Seiberg
duality is the crucial fact that allows this pattern to repeat many times.
For a careful field theoretic discussion of this quasi-periodic RG flow,
see \cite{Strassler}.
If $N= (k+1) M$, where $k$ is an integer, then the 
duality cascade stops after $k$
steps, and we find a $SU(M)\times SU(2M)$ gauge theory. This IR
gauge theory exhibits a multitude of interesting effects visible
in the dual supergravity background, which include the confinement and
chiral symmetry breaking.

\section{Normal Modes of the Warped Throat }

As we shall see below, the $U(1)$ baryonic symmetry of the warped deformed conifold is 
in fact spontaneously broken, since baryonic operators acquire expectation values. 
The corresponding Goldstone boson is a massless pseudoscalar supergravity fluctuation
which has non-trivial monodromy around  
 D-strings at the bottom of the warped deformed conifold \cite{GHK}. Like fundamental strings 
they fall to the bottom of the conifold (corresponding to the IR of the field theory), 
where they have non-vanishing tension. But while F-strings are dual to confining strings, 
D-strings are interpreted as global solitonic strings in the dual cascading 
$SU(M(k+1))\times SU(Mk)$ gauge theory.

Thus the warped deformed conifold naturally incorporates a supergravity description of the supersymmetric Goldstone
mechanism.
Below we review the supergravity dual of a pseudoscalar Goldstone boson, as well as its superpartner, 
a massless scalar glueball \cite{GHK}. In the gauge theory
they correspond to fluctuations in the phase and magnitude of the baryonic condensates, respectively.

\subsection*{The Goldstone mode}

A D1-brane couples to the three-form field strength $F_3$, and therefore we expect a four-dimensional
pseudo-scalar $p(x)$, 
defined so that $*_4 dp = \delta F_3$,
to experience monodromy around the D-string.  

The following ansatz for a linear perturbation of the KS solution \begin{eqnarray} \label{NewPerturbation}
  \delta F^{(3)} &=& \star_4 dp + 
   f_2(\tau)\, dp \wedge dg^5 + f_2'(\tau)\, dp \wedge d\tau \wedge g^5\ , \\
  \delta F^{(5)} &=& (1 + *) \delta F_3 \wedge B_2=
(\star_4 dp
- {\varepsilon^{4/3}\over 6 K^2(\tau)} h(\tau) \,dp\wedge d\tau\wedge g^5) \wedge B_2 \nonumber\ ,
 \end{eqnarray}
where $f_2'= d f_2/d\tau$, falls within the general class of supergravity backgrounds discussed by 
Papadopoulos and Tseytlin \cite{PT}. The metric, dilaton and $B^{(2)}$-field remain unchanged. 
This can be shown to satisfy the linearized supergravity equations \cite{GHK}, 
provided that $d\star_4 dp = 0$, i.e. $p(x)$ is massless,  
and $f_2(\tau)$ satisfies
\bea
-{d\over d\tau} [K^4 \sinh^2 \tau f_2'] &+& {8\over 9 K^2} f_2 \\
&=& { (g_s M \alpha')^2\over 3 \varepsilon^{4/3} } (\tau\coth \tau -1)\left (\coth\tau -
{\tau\over \sinh^2\tau}\right ) \ . \nonumber  
\eea
The normalizable solution of this equation is given by \cite{GHK}
\be
f_2 (\tau) =
   -{2 c \over K^2 \sinh^2 \tau}\int_0^\tau dx \, h(x) \sinh^2 x \,,
\ee
where $c \sim \varepsilon^{4/3}$.
We find that
$f_2\sim \tau$ for small $\tau$, and $f_2\sim \tau e^{-2\tau/3}$ for large $\tau$.

As we have remarked above, the $U(1)$ baryon number symmetry acts as
$A_k \to e^{i\alpha} A_k$, $B_j\to e^{-i\alpha} B_j$.
The massless gauge field in $AdS_5$ dual to the baryon number
current originates from the RR 4-form potential \cite{KW, Ceres}:
\be\label{gaugefield}
\delta C^{(4)}\sim \omega_3\wedge \tilde{A} \ .
\ee 
The zero-mass pseudoscalar glueball arises from the spontaneous
breaking of the global $U(1)_B$ symmetry \cite{Aharony}, as seen from the form of
$\delta F_5$ in (\ref{NewPerturbation}), which contains a term $\sim \omega_3\wedge dp\wedge d\tau$ 
that leads us to identify $\tilde{A} \sim d p$.

If $N$ is an integer multiple of $M$, 
the last step of the cascade leads to a
$SU(2M)\times SU(M)$ gauge theory
coupled to bifundamental fields
$A_i, B_j$  (with $i,j=1,2$). If the $SU(M)$ gauge coupling were turned off, then
we would find an $SU(2M)$ gauge theory coupled to $2M$ flavors. In this $N_f=N_c$ case,
in addition to the usual mesonic branch there exists a baryonic branch of the quantum
moduli space \cite{Seiberg:1994bz}. This is important for the gauge theory
interpretation of the KS background \cite{KS,Aharony}.
Indeed, in addition to mesonic operators
$(N_{ij})^\alpha_\beta \sim (A_i B_j)^{\alpha}_\beta$, the IR gauge theory has 
baryonic operators invariant under the $SU(2M)\times SU(M)$
gauge symmetry, as well as the $SU(2)\times SU(2)$ global symmetry rotating $A_i,B_j$:
\begin{eqnarray} \label{baryonops}
\mathcal{A} &\sim& \epsilon_{\alpha_1\alpha_2\ldots\alpha_{2M}}(A_1)_1^{\alpha_1}(A_1)_2^{\alpha_2}\ldots(A_1)_M^{\alpha_M}
\nonumber\\&&\qquad\qquad\quad(A_2)_1^{\alpha_{M+1}}(A_2)_2^{\alpha_{M+2}}\ldots(A_1)_M^{\alpha_{2M}}\ , \nonumber \\
\mathcal{B} &\sim& \epsilon_{\alpha_1\alpha_2\ldots\alpha_{2M}}(B_1)_1^{\alpha_1}(B_1)_2^{\alpha_2}\ldots(B_1)_M^{\alpha_M}
\nonumber\\&&\qquad\qquad\quad(B_2)_1^{\alpha_{M+1}}(B_2)_2^{\alpha_{M+2}}\ldots(B_1)_M^{\alpha_{2M}}\ .\qquad
\end{eqnarray}
These operators contribute an additional
term to the usual mesonic superpotential:
\begin{equation}
W = \lambda (N_{ij})^\alpha_\beta
(N_{k\ell})_\alpha^\beta\epsilon^{ik}\epsilon^{j\ell}
+ X\left(\det [(N_{ij})^\alpha_\beta]
-{\cal A}{\cal B} - \Lambda_{2M}^{4M}\right) \ ,
\end{equation}
where $X$ can be understood as a Lagrange multiplier.
The supersymmetry-preserving vacua include the baryonic branch:
\be \label{superW}
X = 0 \ ;\quad \ N_{ij} = 0 \ ;\quad 
 {\cal A} {\cal B} = -\Lambda_{2M}^{4M} \ ,
\ee
where the $SO(4)$ global symmetry 
rotating $A_i,B_j$
is unbroken.  In contrast, this global symmetry is broken along the
mesonic branch $N_{ij}\neq 0$. Since the supergravity background of \cite{KS}
is $SO(4)$ symmetric, it is natural to assume that
the dual of this background lies on the baryonic branch of
the cascading theory. 
The expectation values of the baryonic operators
spontaneously break the $U(1)$ baryon number symmetry
$A_k \to e^{i\alpha} A_k$, $B_j\to e^{-i\alpha} B_j$.
The KS background
corresponds to a vacuum where 
$|{\cal A}| = |{\cal B}|=\Lambda_{2M}^{2M}$, which is
invariant under the exchange of the
$A$'s with the $B$'s accompanied by charge conjugation
in both gauge groups. This gives a field theory interpretation to the $\cal{I}$-symmetry of the warped deformed conifold background. As noted in \cite{Aharony}, 
the baryonic branch has complex dimension one, 
and it can be parametrized by $\xi$ as follows
 \be \label{xiDef}
  {\cal A} = i\xi \Lambda_{2M}^{2M} \,,\qquad
  {\cal B} = {i \over \xi} \Lambda_{2M}^{2M} \,.
 \ee
The pseudo-scalar Goldstone mode must correspond to 
changing $\xi$ by a phase, 
since this is precisely what a $U(1)_B$ symmetry transformation does.  

Thus the non-compact warped deformed conifold
exhibits a supergravity
dual of the Goldstone mechanism due to breaking of the global $U(1)_B$ symmetry \cite{Aharony,GHK}. 
On the other hand, if one considered a warped deformed conifold throat embedded in a flux compactification, 
$U(1)_B$ would be gauged, the Goldstone boson $p(x)$ would combine with the $U(1)$ gauge field to 
form a massive vector, and therefore in this situation we would find a manifestation of the 
supersymmetric Higgs mechanism \cite{GHK}.

\subsection*{The Scalar Zero-Mode}
\label{Scalar}

By supersymmetry the massless pseudoscalar is part of a massless ${\mathcal N}=1$
chiral multiplet.  
and therefore there must also be a massless scalar mode 
and corresponding Weyl fermion, with the scalar
corresponding to changing $\xi$ by a positive real factor.  
This scalar zero-mode comes from
a metric perturbation that mixes with the NSNS 2-form
potential.

The warped deformed conifold preserves the
 ${\bf Z}_2$ interchange symmetry ${\cal I}$.
However, the pseudo-scalar mode we found breaks this 
symmetry: 
from the form of the 
perturbations (\ref{NewPerturbation}) we see that 
$\delta F^{(3)}$ is even under the interchange of 
$(\theta_1,\phi_1)$ with $(\theta_2,\phi_2)$,  while $F^{(3)}$
is odd; similarly $\delta F^{(5)}$ is odd while $F^{(5)}$ is even.
Therefore, the scalar mode must also break the ${\cal I}$ symmetry
because in the field theory it breaks the symmetry between
expectation values of $|{\cal A}|$ and of $|{\cal B}|$.
The necessary translationally invariant
perturbation that preserves the $SO(4)$ but breaks the
${\cal I}$ symmetry
is given by the following variation of the NSNS 2-form
and the metric:
\be \label{NSPerturbation}
  \delta B_2 =  \chi(\tau)\, dg^5\ , \; \; \; \; \;
\delta G_{13}  =  \delta G_{24} = 
\lambda (\tau)\ , 
\ee
where, for example $\delta G_{13} = \lambda (\tau)$ means adding 
$2\lambda (\tau)\, g^{(1} g^
{3)}$ to $ds_{10}^2$.
To see that these components of the metric break the ${\cal I}$
symmetry, we note that
\begin{equation}
(e_1)^2 + (e_2)^2 - (\epsilon_1)^2 - (\epsilon_2)^2 =
g^1 g^3 + g^3 g^1+ g^2 g^4+ g^4 g^2\ . 
\end{equation}

Defining
$\lambda (\tau)= h^{1/2} K\sinh (\tau) \, z (\tau)$
one finds \cite{GHK} that all the linearized supergravity equations are
satisfied provided that
\be \label{zdiffeq}
\frac{\left( \left( K \sinh(\tau) \right)^2 z' \right)' }
{(K \sinh(\tau))^2 } = \left(
2 +\frac{8}{9} \frac{1}{K^6}
- \frac{4}{3} \frac{\cosh(\tau)}{K^3}
\right) {z \over \sinh(\tau)^2} \ ,
\ee
and
\be
\chi' =
{1\over 2} g_s M z(\tau) \,{\sinh(2\tau) - 2\tau\over \sinh^2 \tau} \ .
\ee
The solution of (\ref{zdiffeq}) for the zero-mode profile 
is remarkably simple:
\be \label{zresult}
z(\tau)=s\, {
(\tau\coth (\tau)-1) \over
[\sinh(2\tau)-2\tau]^{1/3} }
\ ,
\ee
with 
$s$ a constant.
Like the pseudo-scalar perturbation, the large $\tau$ asymptotic is again 
$z\sim \tau e^{-2\tau/3}$. 
We note that the metric perturbation has the simple form
$\delta G_{13}\sim h^{1/2} [\tau\coth (\tau)-1]
$.
The perturbed metric
$d\tilde s^6_2$ differs from the metric of the deformed
conifold (\ref{metricd}) by
\be
\sim (\tau\coth\tau -1) (g^1 g^3 + g^3 g^1 + g^2 g^4 + g^4 g^2)
\ ,
\ee
which grows as $\ln r$ in the asymptotic radial variable $r$.

The 
scalar zero-mode is actually an exact modulus:
there is a one-parameter family of supersymmetric solutions
which break the ${\cal I}$ symmetry but preserve the $SO(4)$
(an ansatz with these properties was found in \cite{PT}, and
its linearization agrees with (\ref{NSPerturbation})).
These backgrounds, the
 resolved warped deformed conifolds, will be reviewed in the next section.
We add the word resolved because 
both the resolution of the conifold, which is a K\"ahler deformation,
and these resolved warped deformed conifolds
break the ${\cal I}$ symmetry.
In the dual gauge theory
turning on the ${\cal I}$ breaking corresponds to
the transformation
${\cal A} \to (1+s) {\cal A}$,
$\ {\cal B} \to (1+s)^{-1} {\cal B}$ on the baryonic branch.
Therefore, $s$ is dual to the ${\cal I}$ breaking parameter of the 
resolved warped deformed conifold.

The presence of these massless modes is a further indication that the infrared dynamics
of the cascading 
$SU(M(k+1))\times SU(Mk)$ gauge theory, whose supergravity dual is the warped
deformed conifold, is richer than that of
the pure glue ${\cal N}=1$
supersymmetric $SU(M)$ theory.
The former incorporates
a Goldstone supermultiplet, which appears due to the $U(1)_B$ symmetry breaking,
as well as solitonic strings dual to
the D-strings placed at $\tau=0$ in the supergravity
background. 

\subsection*{Massive Glueballs}
 
Let us comment on massive normal modes of the warped deformed conifold. 
These can by found by studying linearized perturbations with four-dimensional momentum $k_\mu$, 
and looking for
the eigenvalues of 
$-k_\mu^2=m_4^2$ at which the resulting equations of motion admit normalizable solutions.
Some early results on the massive glueball spectra were obtained in \cite{Krasnitz, Caceres}.
More recently,
several families of such massive radial excitations which arise from a subset of the deformations 
contained in the PT ansatz were discussed in \cite{BHM}. 
Interestingly, in all cases the mass-squared grows quadratically with the mode number $n$:
\be \label{quadspec}
m_4^2 = A n^2 + \mathcal{O}(n)\ .
\ee
The quadratic dependence on $n$, which is characteristic of Kaluza-Klein theory, is a general
feature
of strongly coupled gauge theories that have weakly curved 10-d gravity duals (see \cite{Karch}
for a discussion). It was also observed that the coefficient $A$ of  the $n^2$ term is 
approximately universal in the sense that the values it takes for different 
towers of excitations are numerically very close to each other \cite{BHM}. 

The towers of massive glueball states based on the pseudoscalar 
Goldstone boson and its scalar superpartner were recently studied in \cite{BDKS}.
The necessary generalization of the scalar ansatz (\ref{NSPerturbation}) to non-zero 4-d momentum is
\bea\label{ansatz}
\delta B^{(2)} &=& \chi(x,\tau)\, dg^5 + \partial_\mu \sigma(x,\tau)\, dx^\mu \wedge g^5 \,, \\
\delta G_{13}& = & \delta G_{24} = \lambda (x,\tau) \,. \eea
A gauge equivalent way of writing (\ref{ansatz}) is
\be
\delta B^{(2)} = (\chi-\sigma)\, dg^5\ -\sigma'\, d\tau \wedge g^5\ .
\ee
Such an ansatz is more general than the generalized PT ansatz used in \cite{BHM} in that
it contains an extra function, $\sigma$.
After some transformations we find that the linearized supergravity equations of motion
reduce to two coupled equations that determine the glueball masses:
\bea
\label{eq1f}
\tilde{z}''-{2\over \sinh^2 \tau}\tilde{z}+
\tilde m^2 { I(\tau)\over K^2(\tau)}
\tilde{z} &=& \tilde m^2 {9 \over 4\cdot 2^{2/3}}\, K(\tau)\, \tilde{w}\ ,
\\
\label{eq2f}
\tilde{w}''-{\cosh^2\tau+1\over \sinh^2 \tau}
\tilde{w}+ \tilde m^2 { I(\tau)\over K^2(\tau)} \tilde{w} &=&
{16\over 9}\, K(\tau)\, \tilde{z} \, ,
\eea
where
\bea
\tilde{z} = h^{-1/2} \lambda(\tau)\ , \qquad
\tilde{w} = { \epsilon^{4/3} \over g_s M \alpha'} K^5\sinh(\tau)^2\sigma'\ ,
\eea
and $\chi$ is determined by the solution for $\sigma$.
Here
the dimensionless eigenvalue $\mt^2$ is related to the mass-squared through
\be
\label{mass}
\tilde m^2 = m_4^2\, {2^{2/3} (g_s M \alpha')^2 \over 6\, \epsilon^{4/3}} \,,
\ee
These coupled equations were solved numerically in \cite{BDKS} yielding radial excitation spectra
of the asymptotic form (\ref{quadspec}) 
with the coefficients of quadratic terms close to those found in
\cite{BHM}.

Note that the scale of these glueball mass-squared $m_4^2$, calculated in the limit
$g_s M\gg 1$, is parametrically
lower than the confining string tension (\ref{conften}).
Using (\ref{mass}) and (\ref{deformedwarp}), we find that
the coefficient $A$ of  the $n^2$ term is
\be 
A \sim {T_s\over g_s M}\ .
\ee
Thus, for radial excitation numbers $n\ll \sqrt{g_s M}$,
these modes are much lighter than the string tension scale, and
therefore much lighter than all glueballs with spin $>2$.
Such anomalously light bound states appear to be special to gauge theories that stay
very strongly coupled in the UV, such as the cascading gauge theory; they do not appear in
asymptotically free gauge theories. Therefore, the anomalously light
low-spin glueballs could perhaps be used as a special ``signature'' of gauge
theories with weakly curved gravity duals, if they are realized in nature.

\section{The Baryonic Branch}
\label{baryonic}

Since the global baryon number symmetry $U(1)_B$ 
is broken by expectation values of baryonic operators, 
the spectrum contains the Goldstone boson found above.
The zero-momentum mode of the scalar superpartner of the Goldstone mode leads to a Lorentz-invariant 
deformation
of the background which describes a small motion
along the baryonic branch. In this section we shall extend the discussion from linearized perturbations around the warped deformed conifold solution to finite deformations, and describe the supergravity backgrounds dual to the complete baryonic branch.
These are the resolved warped deformed conifolds, which preserve the $SO(4)$
global symmetry but break the discrete ${\cal I}$ symmetry of the
warped deformed conifold.

The full set of first-order
equations necessary to describe the entire moduli space of supergravity
backgrounds dual to the baryonic branch, also called the
resolved warped deformed conifolds, was derived and solved numerically in \cite{Butti} (for
a further discussion, see \cite{DKS}).
This continuous family of supergravity solutions is
parameterized by the modulus of $\xi$ (the phase of $\xi$ is not manifest in these backgrounds).
The corresponding metric can be written in the form of the Papadopoulos-Tseytlin
 ansatz \cite{PT}
in the string frame:
\begin{eqnarray}
ds^2 =  h^{-1/2}  \eta_{\mu\nu} dx^\mu dx^\nu
 + e^x ds_\mathcal{M}^2  = h^{-1/2}  dx^2_{1,3} + \sum_{i=1}^6 G_i^2\ ,
\end{eqnarray}
where
\begin{eqnarray} \label{Gforms}
G_1 &\equiv& e^{(x+g)/2}\,e_1\ , \qquad\quad G_3\equiv e^{(x-g)/2}\,(\epsilon_1-a e_1)\ , \\
G_2 &\equiv& {\cosh\tau+a\over \sinh\tau}\,e^{(x+g)/2}\,e_2 + {e^g \over \sinh\tau}\,e^{(x-g)/2}\,(\epsilon_2-a e_2)\ ,\\
G_4 &\equiv& {e^g \over \sinh\tau}\,e^{(x+g)/2}\,e_2 - {\cosh\tau+a\over \sinh\tau}\,e^{(x-g)/2}\,(\epsilon_2-a e_2)\ , \\
G_5 &\equiv& e^{x/2}\,v^{-1/2}d\tau\ , \qquad\
G_6 \equiv e^{x/2}\,v^{-1/2}g^5\ .
\end{eqnarray}

These one-forms describe a basis that rotates as we move along the radial direction, and are particularly convenient since they allow us to write down very simple expressions for the holomorphic $(3,0)$ form 
\begin{equation}
\Omega = (G_1 + i G_2)\wedge (G_3 + i G_4)\wedge (G_5 + i G_6)\ ,
\end{equation}
and the fundamental $(1,1)$ form
\bea
J ={i\over 2} &\Big [&(G_1 + i G_2) \wedge (G_1- i G_2) + (G_3 + i G_4) \wedge (G_3- i G_4) \nonumber \\
&+& (G_5 + i G_6) \wedge (G_5- i G_6) \Big ]\ .
\eea

While in the warped deformed conifold case there was a single warp factor $h(\tau)$, now we
find several additional functions $ x(\tau), g(\tau), a(\tau),
v(\tau)$. The warp factor
$h(\tau)$ is deformed away from (\ref{deformedwarp}) when $|\xi|\neq 1$.

The background also contains the fluxes
\begin{eqnarray}
B^{(2)} &=&
 h_1\,(\epsilon_1 \wedge \epsilon_2 + e_1 \wedge e_2) + \chi\,(e_1 \wedge e_2 - \epsilon_1 \wedge \epsilon_2) \nonumber \\ &&+\, h_2\, (\epsilon_1 \wedge e_2 -  \epsilon_2 \wedge e_1 )
\ , \\
F^{(3)} &=& -{1\over 2} g_5\wedge \big[  \epsilon_1 \wedge \epsilon_2 +  e_1 \wedge e_2 - b\,  (\epsilon_1 \wedge e_2 - \epsilon_2 \wedge e_1) \big] \nonumber \\
&& - {1\over 2}\,  d\tau \wedge \big[ b'\, (\epsilon_1 \wedge e_1 +
\epsilon_2 \wedge e_2) \big]\ ,\\
\tilde{F}^{(5)} &=& {\cal F}^{(5)}  +  *_{10}{\cal F}^{(5)}\ ,\\
{\cal F}^{(5)} &=& -(h_1 + b h_2)\, e_1 \wedge e_2 \wedge \epsilon_1 \wedge \epsilon_2 \wedge \epsilon_3\ , \qquad
\end{eqnarray}
parameterized by functions
$h_1(\tau), h_2(\tau), b(\tau)$ and $\chi(\tau)$. In addition, since
the 3-form flux is not imaginary self-dual for $|\xi|\neq 1$ (i.e.~$\star_6 G_3 \neq i G_3$),
the dilaton $\phi$ now also depends on the
radial coordinate $\tau$. 

The functions $a$ and $v$ satisfy a system of coupled first order differential equations
\cite{Butti} whose solutions are known in closed form only in the warped deformed conifold
and the Chamseddine-Volkov-Maldacena-Nunez (CVMN) \cite{CVMN} limits.
All other functions
$h,x,g,h_1,h_2,b,\chi,\phi$ are unambiguously determined by $a(\tau)$ and $v(\tau)$ through the relations
\begin{eqnarray}\label{hrwdc}
h &=&  \gamma\,U^{-2}\left(e^{-2\phi}-1\right)\ , \qquad \gamma= 2^{10/3} 
(g_s M\alpha')^2\varepsilon^{-8/3} \ ,\\ 
e^{2x} &=& {(bC-1)^2\over 4(aC-1)^2} e^{2g+2\phi}(1-e^{2\phi})\ ,\\
e^{2g} &=& -1-a^2 + 2a\,C\ , \qquad\quad\ b = {\tau\over S}\ ,\\ 
h_2 &=& {e^{2\phi}(bC-1)\over 2S}\ ,\qquad\quad\quad h_1 = -h_2\,C\ ,\\
\chi' &=& a(b-C)(aC-1)\,e^{2(\phi-g)}\ , \\ 
\phi' &=& {\left( C-b\right) {\left(
a\,C-1 \right) }^2 \over \left( b\,C-1 \right) S}\,e^{-2\,g}\ ,
\end{eqnarray}
where $C \equiv -\cosh\tau,\ S \equiv -\sinh\tau$,
and we require $\phi(\infty)=0$. In writing these equations we have 
specialized to the baryonic branch of the cascading gauge theory
by imposing appropriate
boundary conditions at infinity \cite{DKS}, which guarantee that the background asymptotes to the warped conifold solution \cite{KT}.
The full two parameter family of SU(3) structure backgrounds discussed in \cite{Butti} also includes the CVMN solution \cite{CVMN}, which however is characterized by linear dilaton asymptotics that are qualitatively different from the backgrounds discussed here.
The baryonic branch family of supergravity solutions is labelled by one real
``resolution parameter'' $U$ \cite{DKS}.
While the leading asymptotics of all supergravity backgrounds dual to the baryonic
branch are identical to those of the warped deformed conifold, terms subleading at large $\tau$ depend on $U$.
As required, this family of supergravity
solutions preserves the $SU(2)\times SU(2)$ symmetry, but for $U\neq 0$ breaks
the $\mathbb{Z}_2$ symmetry ${\cal I}$.

On the baryonic branch we can consider a transformation that takes
$\xi$ into $\xi^{-1}$, or equivalently
$U$ into $-U$. This transformation leaves $v(\tau)$ invariant and changes $a(\tau)$ as follows
\begin{eqnarray}\label{Utrans}
a\rightarrow -{a\over 1+2a\cosh\tau}\ .
\end{eqnarray}
It is straightforward to check that  $a e^{-g}$ is invariant
while $(1+a\cosh\tau)e^{-g}$ changes sign.
This transformation also exchanges $e^g+a^2e^{-g}$ with $e^{-g}$ and therefore it is equivalent
to the exchange of $(\theta_1,\phi_1)$ and  $(\theta_2,\phi_2)$
involved in the ${\cal I}$-symmetry.

The baryonic condensates have been calculated on the string
theory side of the duality \cite{BDK} 
by identifying the Euclidean D5-branes wrapped over the
deformed conifold, with appropriate gauge fields
turned on, as the appropriate object dual to the baryonic operators in the sense
of gauge/string duality \cite{Aharony}.
Similarly to the case of baryonic operators on the warped resolved conifold discussed in section \ref{resolved}, the field corresponding to a
baryon is recognized as the (semi-classical) equivalent of
$e^{-S_{D5}(r)}$, where $S_{D5}(r)$ is the action of a Euclidean D5-brane
wrapping the Calabi-Yau coordinates up to the radial coordinate
cut-off $r$. The different baryon operators ${\mathcal A},
 {\mathcal B}$, and their conjugates $\overline{{\mathcal A}}, \overline{{\mathcal B}}$, are distinguished by the two possible D5-brane orientations, and the
two possible $\kappa$-symmetric choices for the world volume gauge
field that has to be turned on inside the D5-brane.

This identification is legitimate since in the cascading theory, which is
near-AdS in the UV, equation (\ref{sourceVEV}) holds modulo powers of $\ln r$
\cite{Aharony:2005zr,Aharony:2006ce}.
Due to the absence of sources for baryonic operators we again have $\varphi_0=0$, and 
$e^{-S_{D5}(r)}$ gives the dimensions of the baryon operators, and the values of their
condensates.

In contrast to the simpler case of the warped resolved conifold, a world-volume gauge bundle $F^{(2)}=dA^{(1)}$ is required by $\kappa$-symmetry in this case, which leads to the conditions \cite{Marino:1999af} that
$\mathcal{F} \equiv F^{(2)}+ B^{(2)}$ be a $(1,1)$-form, and that
\begin{equation}\label{GeomKappaCond}
{1\over2!}J\wedge J \wedge \mathcal{F} - {1\over3!} \mathcal{F} \wedge \mathcal{F} \wedge \mathcal{F} = \mathfrak{g} \left({1\over3!}J \wedge J \wedge J - {1\over2!}J\wedge \mathcal{F} \wedge \mathcal{F} \right).
\end{equation}
Here $\mathfrak{g}$ would simply be a constant if the internal manifold were Calabi-Yau, but 
since we are dealing with a generalized Calabi-Yau
with fluxes, $\mathfrak{g}$ becomes coordinate dependent, a function of $\tau$ in our case.

The
$SU(2)\times SU(2)$ invariant ansatz for the gauge potential is given by
\begin{equation} \label{gaugeans}
A^{(1)} = \zeta(\tau) g^5\ ,
\end{equation}
which together with (\ref{GeomKappaCond}) implies that $\zeta$ has to satisfy the differential equation
\begin{equation}
\zeta' = {e^x(\mathfrak{g} \mathfrak{a} + \mathfrak{b})
\over v (\mathfrak{a} - \mathfrak{g} \mathfrak{b})}\ ,
\end{equation}
where we have defined
\begin{eqnarray} \label{aandb}
\mathfrak{a}(\zeta,\tau) &\equiv& e^{-2x}[e^{2x} + h_2^2 \sinh^2(\tau)-(\zeta+\chi)^2]\ ,\nonumber \\
\mathfrak{b}(\zeta,\tau) &\equiv& 2 e^{-x-g}\sinh(\tau) [a(\zeta+\chi) - h_2(1+a\cosh(\tau))]\ .
\end{eqnarray}

Using the explicitly known Killing spinors of the baryonic branch backgrounds (or from an equivalent argument starting from the Dirac-Born-Infeld equations of motion) one can show \cite{BDK} that  
\begin{equation} \label{g-function}
\mathfrak{g} = - {e^{-x+g} h_2\sinh(\tau) \over (1+a\cosh(\tau))} = {e^{\phi}\over\sqrt{1-e^{2\phi}}}\ ,
\end{equation}
which determines $\zeta$ and thus the action of the Euclidean D5-brane:
\be \label{D5action}
S_{D5} \sim \int  d \tau\, e^{-\phi}\sqrt{\det{G+\mathcal{F}}} = \int  d \tau\,
{e^{-\phi}e^{3x}\sqrt{1 + \mathfrak{g}^2}\,(\mathfrak{a}^2+\mathfrak{b}^2)
\over v|\mathfrak{a} - \mathfrak{g} \mathfrak{b}|}\ .
\ee

The dimension of the baryon
operators in the KS background can be extracted from the divergent terms of the D5-brane action as a function
of the radial cut-off $r$, which leads to
\begin{equation} \label{bardim}
\Delta(r) = r {d S_{D5}(r) \over d r}  =  {27 g_s^2 M^3 \over 16 \pi^2} (\ln(r))^2 + \mathcal{O}(\ln(r))\ .
\end{equation}
To compare this with the cascading gauge theory, we use the fact that the 
baryon operators of the $SU(M(k+1))\times SU(Mk)$
theory have the schematic
form $\mathcal{A} \sim (A_1A_2)^{k(k+1)M/2}$ and $\mathcal{B}\sim (B_1B_2)^{k(k+1)M/2}$,
with appropriate contractions described in \cite{Aharony}.
For large $k$, their dimensions are $\Delta(k) \approx 3 M k(k+1)/4$. 
If we remember that the radius at which the $k$-th Seiberg duality is performed is given by
\begin{equation}
r(k) \sim \varepsilon^{2/3} \exp\left({2\pi k \over 3 g_s M}\right)\ ,
\end{equation}
we find that the leading term in the operator dimension agrees with (\ref{bardim}).

The baryon expectation
values as a function of $U$ can be computed by evaluating the finite terms in the action (\ref{D5action}). The baryonic condensates calculated in this fashion \cite{BDK} satisfy the important condition that $\langle \mathcal{A} \rangle \langle \mathcal{B} \rangle =$ const. along the whole baryonic branch.
This leads to a precise relation between the baryonic branch
modulus $|\xi|$ in the gauge theory (\ref{xiDef}) and the modulus $U$ in the
dual supergravity description. 

Furthermore, pseudoscalar perturbations around the warped deformed conifold background are seen explicitly to shift the phase
of the baryon expectation value, through the Chern-Simons term in the D5-brane action, as required for consistency.

\section{Cosmology in the Throat}

A promising framework for realizing cosmological inflation \cite{Inflation} in
string theory is D-brane inflation (for reviews and more complete references, see
\cite{reviews}).
The original proposal \cite{Dvali} was to consider 
a D3-brane and a $\overline {\rm D3}$-brane separated by some distance along the compactified
dimensions, and with their world volumes spanning the 4 observable coordinates $x^\mu$. 
From the 4-d point of view, the distance $r$ between the branes is a scalar field that is identified
with the inflaton. However, in flat space the Coulomb potential $\sim {1/ r^4}$ is typically too steep
to support slow-roll inflation. An ingenious proposal to circumvent this problem
\cite{KKLMMT} is to place the brane-antibrane pair in a warped throat region of a flux compactification,
of which the warped deformed conifold is an explicitly known and ubiquitous \cite{Hebecker} example.

The $\overline {\rm D3}$-brane breaks supersymmetry and experiences potential a 
${2T_3 / h(\tau) }$ attracting it to the bottom of the conifold, $\tau=0$. 
Its energy density there, ${2T_3 / h(0) }$,
plays the important role of ``uplifting'' the negative 4-d cosmological constant
to a positive value in the KKLT model for moduli stabilization \cite{KKLT}.
When a mobile D3-brane is added, it perturbs the background warp factor.
The energy density of the $\overline {\rm D3}$-brane becomes 
\begin{equation} 
V(r)= {2 T_3\over h(0)+ \delta h(0,r)} \approx  
{2T_3\over h(0) } -2T_3 {\delta h(0,r)\over h(0)^2}
\ ,\end{equation}
where $\delta h(0,r)$ is the perturbation of the warped factor at the position of the $\overline {\rm D3}$-brane $r=0$, caused by the D3-brane at radial coordinate $r$.
For a D3-brane far from the tip of the throat, with radial coordinate $r\gg \varepsilon^{2/3}$,
$\delta h(0,r)\approx {27 / (32 \pi^2 T_3 r^4)}$ \cite{BDKMMM}.
Thus, the potential assumes the form \cite{KKLMMT, Baumann:2007ah}
(note that the definition of $r^2$ here differs by a factor of $3/2$ from that in
\cite{BDKMMM})
\begin{equation} \label{antib}
V(r) =  {2T_3\over h(0) }
- \frac{27 }{16 \pi^2 h(0)^2 r^4}\ .
\end{equation}
Thus, the force on the D3-brane is suppressed by
a small factor $h(0)^{-2}$ compared to the force
in unwarped space used in the original model \cite{Dvali}.
We recall that
\be \label{warptip}
h(0) = a_0 (g_s M \alpha')^2 2^{2/3} \varepsilon^{-8/3} \ ,\qquad a_0 \approx 0.71805\ .
\ee
In flux compactifications containing a long KS throat, $h(0)$ is of order
$e^{8\pi K/3g_s M}$ \cite{GKP}.
The flattening of the brane-antibrane potential by exponential warping is an important factor in
constructing realistic brane inflation scenarios.

There are possibilities other than 
a $\overline {\rm D3}$-brane at the bottom of the throat for creating a slow-varying potential for
the mobile D3-brane. As pointed out in \cite{DKS}, if the throat is taken to be a
{\it resolved} warped deformed conifold, which was reviewed in section \ref{baryonic}, then the potential 
experienced by the D3-brane is
\be
V(\tau)= T_3
h^{-1}(\tau) (e^{-\phi(\tau)} -1)
\ .
\ee
The first term comes from
the Born-Infeld term and has a factor of $e^{-\phi(\tau)}$;
the second term, originating from
the interaction with the background 4-form $C_{0123}$, does not
have this factor. For the KS solution ($U=0$), $\phi(\tau)=0$;
therefore, the potential vanishes and the
D3-brane may be located at any point on the deformed conifold.
For
$U\neq 0$ we may use (\ref{hrwdc}) to write
\be
V(\tau)= {T_3\over
\gamma} {U^2\over  e^{-\phi(\tau)} +1}
\ .
\ee
Since $\phi(\tau)$ is a monotonically increasing function,
the D3-brane is attracted to $\tau=0$.

For large enough $\tau$, the potential becomes
\be \label{resolpot}
V(\tau)= {T_3\over
\gamma} \left [{U^2\over 2} - {3 U^4\over 256} (4\tau-1) e^{-4\tau/3} + \ldots \right ]
\ .
\ee
Since $e^{-4\tau/3}\sim \varepsilon^{8/3} r^{-4}$, this is rather similar to the 
brane-antibrane potential (\ref{antib}), and has the additional feature that the resolution
parameter $U$ may be varied. In a complete treatment, $U$ should be determined
by the details of the compactification.

\subsection*{Cosmic Strings}

In addition to suppressing the D3-brane potential,
the large value of $h(0)$ is responsible for the viability of cosmic strings
in flux compactifications containing long warped throats. Cosmic strings with Planckian
tensions are ruled out by the CMB spectrum and other astrophysical observations
(for a review, see \cite{JPolchinski}). The current constraints on cosmic string tension $\mu$
suggest $G \mu \ll 10^{-7}$, and they continue to improve. In traditional models
where the string scale is not far from the 4-d Planck scale,
the tension of a fundamental string far from the warped throat, ${1 / (2\pi \alpha')}$, badly violates
this constraint. However, at the bottom of the throat the tension is reduced to
${1 / (2\pi \alpha' \sqrt{h(0)})}$ which can be consistent with the constraint \cite{Copeland}. 
It is remarkable that such a cosmic string has a dual description as a confining string 
in the cascading gauge theory dual to the throat. This shows that the phenomenon of color confinement
may have implications reaching far beyond the physics of hadrons. 

Type IIB flux compactifications with long warped throats may contain
a variety of species of cosmic strings. $q$ fundamental strings at the bottom of the throat
may form a bound state, which is described by a D3-brane wrapping a 2-sphere within
the 3-sphere \cite{HK}.
A D-string at the bottom of the throat is dual to
a certain solitonic string in the gauge theory \cite{GHK}. 
Furthermore, $p$ D-strings and $q$ F-strings can
bind into a $(p,q)$ string \cite{Firouzjahi}. 
Networks of such $(p,q)$ cosmic strings have various characteristic
features in their evolution and interaction probabilities which could distinguish them from other cosmic string models.
Importantly, such strings are copiously produced during the brane-antibrane annihilation that
follows the brane inflation \cite{Sarangi}. However, in models where some D3-branes remain at the bottom
of the throat after inflation (for example, \cite{DKS}), long cosmic strings cannot exist because they break on
the D-branes \cite{JPolchinski}. Thus, non-observation of cosmic strings would not rule out
D-brane inflation.

\subsection*{Compactification Effects}

The potentials (\ref{antib}), (\ref{resolpot}) are very flat for large $r$, approaching a constant
that arises due to D-term breaking of supersymmetry. However, additional contributions to the potential
that arise due to moduli stabilization effects tend to destroy this flatness and generally render
slow-roll inflation impossible.
Indeed, in the compactified setting, the contribution of the brane-antibrane interaction to
the 4-d ``Einstein frame'' potential is
\be \label{equ:Dterm} V_D(\rho, r)= \frac{V(r)}{U^2(\rho,r)}\, ,
\ee
The extra factor comes from the DeWolfe-Giddings K\"ahler potential \cite{DeWG} which depends
both on 
the volume modulus, $\rho$, and the 
D3-brane position $z_\alpha$, $\alpha=1,2,3$:
\be \label{equ:KKK} \kappa^2
{\cal K}(\rho,\overline{\rho},z_\alpha,\overline{z_\alpha})  =
- 3 \log [\rho + \bar \rho - \beta k(z_\alpha, \overline{z_\alpha}) ]
\equiv - 3 \log U \, .
\ee
Here $k(z_\alpha, \overline{z_\alpha})$ denotes the K\"ahler
potential of the Calabi-Yau manifold, which in the throat reduces to
\be \label{Kpot}
k = \frac{3}{2} \left( 
\sum_{i=1}^4 |z_i|^2 \right)^{2/3} = r^2 \ , \ee
where we ignored the deformation $\varepsilon$.
The normalization constant $\beta$ in (\ref{equ:KKK}) may be expressed as
\be
\label{equ:gamma}
\beta \equiv \frac{\sigma_0}{3} \frac{T_3}{M_P^2}\, ,
\ee
where
$2 \sigma_0 \equiv 2 \sigma_\star(0) = \rho_\star(0) + \bar \rho_\star(0)$
is the stabilized value of the K\"ahler modulus
when the D3-brane is near the tip of the throat. For a D3-brane far from the tip,
we may ignore the deformation of
the conifold and use
\be 
\label{equ:U}
U(\rho, r)= \rho
+\overline{\rho} - \beta r^2 \ .
\ee
The $r$-dependence from the factor $U^{-2}(\rho,r)$ spoils the flatness of the
potential even in the region where $V(r)$ is very flat.

The complete inflaton potential 
\be \label{equ:Vtot} V_{tot} = V_F(\rho,z_\alpha) + V_D(\rho,r) \,  \ee
also includes the F-term contribution whose standard expression in
${\cal N}=1$ supergravity is
\be
\label{equ:VF} V_F = e^{\kappa^2 {\cal K}} \Bigl[ D_\Sigma W
{\cal K}^{\Sigma \overline{\Omega}} \overline{D_\Omega W}- 3
\kappa^2 W \overline{W}\Bigr]\, , \quad \quad \kappa^2 =
M_P^{-2} \equiv 8 \pi G \, ,\ee 
where $\{Z^\Sigma\} \equiv
\{\rho, z_\alpha; \alpha=1,2,3 \}$ and $D_\Sigma W = \partial_\Sigma W + 
\kappa^2 (\partial_\Sigma {\cal K}) W$. 
The
superpotential $W$ has the structure
\be
\label{equ:W}
 W(\rho,z_\alpha) = W_0 + A(z_\alpha) e^{- a \rho}\, , \quad \quad \quad
a \equiv \frac{2 \pi}{n}\, .
\ee
where the second, nonperturbative term
arises either from strong gauge dynamics on a stack of $n>1$ D7-branes
or from Euclidean D3-branes (with $n =1$ ).  We assume that either sort of brane
supersymmetrically wraps a four-cycle in the warped throat that
is specified by a holomorphic embedding equation $f(z_\alpha)=0$.
The warped volume of the four-cycle governs the magnitude of the nonperturbative
effect, by affecting the gauge coupling on the D7-branes
(equivalently, the action of Euclidean D3-branes) wrapping this four-cycle.
The presence of a D3-brane gives rise to a perturbation to the warp factor, and
this leads to a correction to the warped four-cycle volume.
This correction depends on the D3-brane position and is responsible 
for the prefactor $A(z_\alpha)$ \cite{GM}.
In \cite{BDKMMM}, 
D3-brane backreaction on the warped four-cycle volume was calculated leading to
a simple formula
\be \label{equ:A} 
A(z_\alpha) = A_0 \left(\frac{f(z_\alpha)}{f(0)} \right)^{1/n}\, . 
\ee

The canonical inflaton $\varphi$ is proportional to $r$, the radial location of the D3-brane.
Using (\ref{equ:Vtot})
to compute the slow-roll parameter \be  \eta \equiv M_P^2 \frac{V_{,\varphi\varphi}}{V} \, , \ee
one finds \be
\label{equ:eta}
 \eta = \frac{2}{3} + \Delta \eta(\varphi) \, , \ee
where the $2/3$ arises from the K\" ahler potential
(\ref{equ:KKK}), (\ref{equ:U}), and $\Delta \eta$ from the variation of $A(z_\alpha)$.
A simple model studied in \cite{Baumann:2007ah} was based on the Kuperstein embedding of
a stack of D7-branes
\be
f(z_1)= \mu- z_1 =0 
\ .
\ee
Explicit calculation of the full inflaton potential in this model 
\cite{Baumann:2007ah,Krause} shows that it is possible
to fine-tune the parameters to achieve an inflection point in the vicinity of which 
slow-roll inflation is possible.

Other supergravity and string theory constructions where such Inflection Point Inflation
may be achieved were proposed in \cite{Holman,MSSM,Itzhaki,Linde}. While  
such models are fine-tuned, and the initial conditions have to be chosen carefully
to prevent the field from running through the inflection point with high speed (see, however,
\cite{Itzhaki,Underwood} for possible ways to circumvent this problem) the model is fairly predictive.
In particular, the spectral index $n_s$ is around $0.93$ in the limit that the total number of
e-folds is large during inflation. This fact may render this model distinguishable from
others by more precise observations of the CMB.

\section{Summary}

Throughout its history, string theory has been intertwined with 
the theory of strong interactions. The AdS/CFT 
correspondence\cite{jthroat,US,EW} has succeeded
in making precise connections between conformal 4-dimensional
gauge theories and
superstring theories in 10 dimensions. This duality leads to a multitude
of dynamical predictions about strongly coupled gauge theories. 
While many of these predictions are difficult to check, recent applications of
methods of exact integrability to planar ${\cal N}=4$ SYM theory have produced some
impressive tests of the corresponence for operators with high spin.
When extended to theories at finite temperature, the correspondence serves
as a theoretical laboratory for studying a novel state of matter:
a gluonic plasma at very strong coupling. This appears to have
surprising connections to the new state of matter, sQGP, which was observed at 
RHIC and will be further studied at the LHC.

Breaking symmetries in the AdS/CFT correspondence is important for bringing it closer
to the real world. Some of the supersymmetry may be broken by considering 
D3-branes at conical singularities; the case of the conifold is discussed
in detail in these lectures. In this set-up, breaking of gauge symmetry 
typically leads to a resolution of the 
singularity. The associated breaking of global symmetry leads to the appearance of Goldstone bosons and
global strings.
Extensions of the gauge-string duality to confining gauge theories
provide new geometrical viewpoints on such important phenomena as
chiral symmetry breaking and dimensional transmutation, which are encoded in the dual
smooth warped throat background. Embedding of the throat into flux compactifications of string theory
allows for an interesting interplay between gauge-string duality and models of particle physics and
cosmology. For example, D3-branes rolling in the throat might model inflation while various 
strings attracted to the bottom of the throat may describe cosmic strings.
All of this raises hopes that the new window into strongly coupled gauge theory
opened by the discovery of gauge-string dualities will one day lead to new striking connections
between string theory and the real world.

\section*{Acknowledgments}
We thank the organizers of the Les Houches 2007 summer school ``String Theory and the Real World''
for hospitality and for organizing
a very stimulating school.
We are very grateful to 
D. Baumann, S. Benvenuti, A. Dymarsky, 
S. Gubser, C. Herzog, J. Maldacena, L. McAllister, A. Murugan, A. Peet, A. Polyakov,
D. Rodriguez-Gomez,
A. Scardicchio, A. Solovyov, M. Strassler, A. Tseytlin, J. Ward and E. Witten
 for collaboration on some of the papers reviewed here.
We also thank D. Baumann, C. Herzog and A. Murugan for helpful comments on the manuscript.
This research is
supported in part by the National Science Foundation 
Grant No.~PHY-0243680.


\begin{thebibliography}{99}

\bibitem{jthroat}
  J.~M.~Maldacena,
  ``The large N limit of superconformal field theories and supergravity,''
  Adv.\ Theor.\ Math.\ Phys.\  {\bf 2}, 231 (1998)
  [arXiv:hep-th/9711200].
  %%CITATION = HEP-TH 9711200;%%

\bibitem{US}
  S.~S.~Gubser, I.~R.~Klebanov and A.~M.~Polyakov,
  ``Gauge theory correlators from non-critical string theory,''
  Phys.\ Lett.\ B {\bf 428}, 105 (1998)
  [arXiv:hep-th/9802109].
  %%CITATION = HEP-TH 9802109;%%

\bibitem{EW}
E.~Witten,
  ``Anti-de Sitter space and holography,''
Adv.\ Theor.\ Math.\ Phys.\  {\bf 2}, 253 (1998)
  [arXiv:hep-th/9802150].
  %%CITATION = HEP-TH 9802150;%%

\bibitem{qcdstring}
  I.~R.~Klebanov,
  ``QCD and string theory,''
  Int.\ J.\ Mod.\ Phys.\  A {\bf 21}, 1831 (2006)
  [arXiv:hep-ph/0509087].
  %%CITATION = IMPAE,A21,1831;%%

\bibitem{Dolen}
  R.~Dolen, D.~Horn and C.~Schmid,
``Finite Energy Sum Rules And Their Application To Pi N Charge Exchange,''
  Phys.\ Rev.\  {\bf 166}, 1768 (1968).
  %%CITATION = PHRVA,166,1768;%%


\bibitem{Veneziano}
  G.~Veneziano,
  ``Construction Of A Crossing - Symmetric, Regge Behaved Amplitude For
  Linearly Rising Trajectories,''
  Nuovo Cim.\ A {\bf 57}, 190 (1968).
  %%CITATION = NUCIA,A57,190;%%
  
  
\bibitem{Nambu}
Y. Nambu, 
``Quark model and the factorization of the Veneziano amplitude,''
in {\em Symmetries and Quark Models}, ed. R. Chand, Gordon and Breach
(1970).
  
\bibitem{Nielsen}
H. B. Nielsen, 
``An almost physical interpretation of the integrand
of the n-point Veneziano amplitude,'' 
submitted to the 15th 
International Conference on High Energy Physics, Kiev (1970).

\bibitem{Susskind}
L. Susskind,  
``Dual-symmetric theory of hadrons,''
{\em Nuovo Cim.} {\bf 69A} (1970) 457.

\bibitem{DiVecchia}
  P.~Di Vecchia and A.~Schwimmer,
  ``The beginning of string theory: a historical sketch,''
  arXiv:0708.3940 [physics.hist-ph];
  %%CITATION = ARXIV:0708.3940;%%
P.~Di Vecchia,
  ``The birth of string theory,''
  arXiv:0704.0101 [hep-th].
  %%CITATION = ARXIV:0704.0101;%%

\bibitem{GWP}
  D.~J.~Gross and F.~Wilczek,
  ``Ultraviolet Behavior Of Non-Abelian Gauge Theories,''
  Phys.\ Rev.\ Lett.\  {\bf 30}, 1343 (1973);
  %%CITATION = PRLTA,30,1343;%%
  H.~D.~Politzer,
  ``Reliable Perturbative Results For Strong Interactions?,''
  Phys.\ Rev.\ Lett.\  {\bf 30}, 1346 (1973).
  %%CITATION = PRLTA,30,1346;%%

\bibitem{Scherk}
J. Scherk and J. Schwarz, ``Dual models for non-hadrons,''
{\em Nucl. Phys.} {\bf B81} (1974) 118.

\bibitem{Yoneya}
  T.~Yoneya,
  ``Connection of Dual Models to Electrodynamics and Gravidynamics,''
  Prog.\ Theor.\ Phys.\  {\bf 51}, 1907 (1974).
  %%CITATION = PTPKA,51,1907;%%

\bibitem{Wilson}
  K.~G.~Wilson,
  ``Confinement Of Quarks,''
  Phys.\ Rev.\ D {\bf 10}, 2445 (1974).
  %%CITATION = PHRVA,D10,2445;%%

\bibitem{Nambunew}
  Y.~Nambu,
  ``QCD And The String Model,''
  Phys.\ Lett.\ B {\bf 80}, 372 (1979).
  %%CITATION = PHLTA,B80,372;%%

\bibitem{Andersson}
  B.~Andersson, G.~Gustafson, G.~Ingelman and T.~Sjostrand,
  ``Parton Fragmentation And String Dynamics,''
  Phys.\ Rept.\  {\bf 97}, 31 (1983).
  %%CITATION = PRPLC,97,31;%%

\bibitem{Luscher:1980fr}
  M.~Luscher, K.~Symanzik and P.~Weisz,
  ``Anomalies Of The Free Loop Wave Equation In The Wkb Approximation,''
  Nucl.\ Phys.\ B {\bf 173}, 365 (1980).
  %%CITATION = NUPHA,B173,365;%%

\bibitem{Luscher}
  M.~Luscher and P.~Weisz,
  ``Quark confinement and the bosonic string,''
  JHEP {\bf 0207}, 049 (2002)
  [arXiv:hep-lat/0207003].
  %%CITATION = HEP-LAT 0207003;%%

\bibitem{GT}
  G.~'t Hooft,
  ``A Planar Diagram Theory For Strong Interactions,''
  Nucl.\ Phys.\ B {\bf 72}, 461 (1974).
  %%CITATION = NUPHA,B72,461;%%

\bibitem{Polyakov}
  A.~M.~Polyakov,
``Quantum Geometry Of Bosonic Strings,''
  Phys.\ Lett.\ B {\bf 103}, 207 (1981); 
  %%CITATION = PHLTA,B103,207;%%

\bibitem{Polyakovnew}
 A.~M.~Polyakov,
``String theory and quark confinement,''
  Nucl.\ Phys.\ Proc.\ Suppl.\  {\bf 68}, 1 (1998)
  [arXiv:hep-th/9711002].
  %%CITATION = HEP-TH 9711002;%%

\bibitem{Polch}
  J.~Polchinski,
  ``Dirichlet-Branes and Ramond-Ramond Charges,''
  Phys.\ Rev.\ Lett.\  {\bf 75}, 4724 (1995)
  [arXiv:hep-th/9510017].
  %%CITATION = HEP-TH 9510017;%%

\bibitem{Horowitz}
  G.~T.~Horowitz and A.~Strominger,
  ``Black strings and P-branes,''
  Nucl.\ Phys.\ B {\bf 360}, 197 (1991).
  %%CITATION = NUPHA,B360,197;%%

\bibitem{absorption}
I.~R.~Klebanov,
  ``World-volume approach to absorption by non-dilatonic branes,''
  Nucl.\ Phys.\ B {\bf 496}, 231 (1997)
  [arXiv:hep-th/9702076]; 
  %%CITATION = HEP-TH 9702076;%%
S.~S.~Gubser, I.~R.~Klebanov and A.~A.~Tseytlin,
  ``String theory and classical absorption by three-branes,''
  Nucl.\ Phys.\ B {\bf 499}, 217 (1997)
  [arXiv:hep-th/9703040]; 
  %%CITATION = HEP-TH 9703040;%%
S.~S.~Gubser and I.~R.~Klebanov,
  ``Absorption by branes and Schwinger terms in the world volume theory,''
  Phys.\ Lett.\ B {\bf 413}, 41 (1997)
  [arXiv:hep-th/9708005].
  %%CITATION = HEP-TH 9708005;%%

\bibitem{Malda}
  J.~M.~Maldacena,
  ``Wilson loops in large N field theories,''
  Phys.\ Rev.\ Lett.\  {\bf 80}, 4859 (1998)
  [arXiv:hep-th/9803002]; \\
  %%CITATION = HEP-TH 9803002;%%
S.~J.~Rey and J.~T.~Yee,
  ``Macroscopic strings as heavy quarks in large N gauge theory and  anti-de
  Sitter supergravity,''
  Eur.\ Phys.\ J.\ C {\bf 22}, 379 (2001)
  [arXiv:hep-th/9803001].
  %%CITATION = HEP-TH 9803001;%%

\bibitem{Berenstein}
  D.~Berenstein, J.~M.~Maldacena and H.~Nastase,
  ``Strings in flat space and pp waves from N = 4 super Yang Mills,''
  JHEP {\bf 0204}, 013 (2002)
  [arXiv:hep-th/0202021].
  %%CITATION = HEP-TH 0202021;%%

\bibitem{Gubser:2002tv}
  S.~S.~Gubser, I.~R.~Klebanov and A.~M.~Polyakov,
  ``A semi-classical limit of the gauge/string correspondence,''
  Nucl.\ Phys.\ B {\bf 636}, 99 (2002)
  [arXiv:hep-th/0204051].
  %%CITATION = HEP-TH 0204051;%%

\bibitem{Tseytlinrev}
  A.~A.~Tseytlin,
  ``Semiclassical strings and AdS/CFT,''
  arXiv:hep-th/0409296.
  %%CITATION = HEP-TH 0409296;%%

\bibitem{Beisert}
%\cite{Belitsky:2004cz}
%\bibitem{Belitsky:2004cz}
  A.~V.~Belitsky, V.~M.~Braun, A.~S.~Gorsky and G.~P.~Korchemsky,
  ``Integrability in QCD and beyond,''
  Int.\ J.\ Mod.\ Phys.\ A {\bf 19}, 4715 (2004)
  [arXiv:hep-th/0407232];
  %%CITATION = HEP-TH 0407232;%%
N.~Beisert,
  ``The dilatation operator of N = 4 super Yang-Mills theory and
  integrability,''
  Phys.\ Rept.\  {\bf 405}, 1 (2005)
  [arXiv:hep-th/0407277];
  %%CITATION = HEP-TH 0407277;%%


\bibitem{GrossWilczek}
  D.~J.~Gross and F.~Wilczek,
  ``Asymptotically Free Gauge Theories. 2,''
  Phys.\ Rev.\ D {\bf 9}, 980 (1974);
  %%CITATION = PHRVA,D9,980;%%
H.~Georgi and H.~D.~Politzer,
  ``Electroproduction Scaling In An Asymptotically Free Theory Of Strong Interactions,''
  Phys.\ Rev.\ D {\bf 9}, 416 (1974).
  %%CITATION = PHRVA,D9,416;%%

\bibitem{Korchemsky}
  G.~P.~Korchemsky,
  ``Asymptotics Of The Altarelli-Parisi-Lipatov Evolution Kernels Of Parton Distributions,''
  Mod.\ Phys.\ Lett.\ A {\bf 4}, 1257 (1989);
  %%CITATION = MPLAE,A4,1257;%%
G.~P.~Korchemsky and G.~Marchesini,
  ``Structure function for large x and renormalization of Wilson loop,''
  Nucl.\ Phys.\ B {\bf 406}, 225 (1993)
  [arXiv:hep-ph/9210281].
  %%CITATION = HEP-PH 9210281;%%


\bibitem{Floratos}
E.G. Floratos, D.A. Ross, Christopher T. Sachrajda, ``Higher Order Effects In Asymptotically Free Gauge Theories. 2. Flavor
  Singlet Wilson Operators And Coefficient Functions,'' {\em Nucl.
Phys.} {\bf B152} (1979) 493.

\bibitem{Sterman}
  G.~Sterman and M.~E.~Tejeda-Yeomans,
  ``Multi-loop amplitudes and resummation,''
  Phys.\ Lett.\ B {\bf 552}, 48 (2003)
  [arXiv:hep-ph/0210130].
  %%CITATION = HEP-PH 0210130;%%

\bibitem{FT}
  S.~Frolov and A.~A.~Tseytlin,
  ``Semiclassical quantization of rotating superstring in AdS(5) x S(5),''
  JHEP {\bf 0206}, 007 (2002)
  [arXiv:hep-th/0204226].
  %%CITATION = HEP-TH 0204226;%%

\bibitem{Belitsky:2006en}
  A.~V.~Belitsky, A.~S.~Gorsky and G.~P.~Korchemsky,
  ``Logarithmic scaling in gauge / string correspondence,''
  Nucl.\ Phys.\ B {\bf 748}, 24 (2006)
  [arXiv:hep-th/0601112].
  %%CITATION = HEP-TH 0601112;%%

\bibitem{Kotikov}
  A.~V.~Kotikov, L.~N.~Lipatov, A.~I.~Onishchenko and V.~N.~Velizhanin,
  ``Three-loop universal anomalous dimension of the Wilson operators in N =  4
  SUSY Yang-Mills model,''
  Phys.\ Lett.\ B {\bf 595}, 521 (2004)
  [arXiv:hep-th/0404092].
  %%CITATION = HEP-TH 0404092;%%

\bibitem{Bern}
  Z.~Bern, L.~J.~Dixon and V.~A.~Smirnov,
  ``Iteration of planar amplitudes in maximally supersymmetric Yang-Mills
  theory at three loops and beyond,''
  arXiv:hep-th/0505205.
  %%CITATION = HEP-TH 0505205;%%

\bibitem{Moch}
  S.~Moch, J.~A.~M.~Vermaseren and A.~Vogt,
  ``The three-loop splitting functions in QCD: The non-singlet case,''
  Nucl.\ Phys.\ B {\bf 688}, 101 (2004)
  [arXiv:hep-ph/0403192].
  %%CITATION = HEP-PH 0403192;%%

\bibitem{Kruczenski}
  M.~Kruczenski,
  ``A note on twist two operators in N = 4 SYM and Wilson loops in Minkowski
  signature,''
  JHEP {\bf 0212}, 024 (2002)
  [arXiv:hep-th/0210115].
  %%CITATION = HEP-TH 0210115;%%

\bibitem{Minahan}
  J.~A.~Minahan and K.~Zarembo,
  ``The Bethe-ansatz for N = 4 super Yang-Mills,''
  JHEP {\bf 0303}, 013 (2003)
  [arXiv:hep-th/0212208];
  %%CITATION = HEP-TH 0212208;%%
N.~Beisert, C.~Kristjansen and M.~Staudacher,
  ``The dilatation operator of N = 4 super Yang-Mills theory,''
  Nucl.\ Phys.\ B {\bf 664}, 131 (2003)
  [arXiv:hep-th/0303060];
  %%CITATION = HEP-TH 0303060;%%
N.~Beisert and M.~Staudacher,
  ``The N = 4 SYM integrable super spin chain,''
  Nucl.\ Phys.\ B {\bf 670}, 439 (2003)
  [arXiv:hep-th/0307042].
  %%CITATION = HEP-TH 0307042;%%

\bibitem{Eden}
  B.~Eden and M.~Staudacher,
  ``Integrability and transcendentality,''
J.\ Stat.\ Mech.\  {\bf 0611}, P014 (2006)
  [arXiv:hep-th/0603157].
  %%CITATION = JSTAT,0611,P014;%%

\bibitem{BHL}
  N.~Beisert, R.~Hernandez and E.~Lopez,
  ``A crossing-symmetric phase for AdS(5) x S**5 strings,''
JHEP {\bf 0611}, 070 (2006)
  [arXiv:hep-th/0609044].
  %%CITATION = JHEPA,0611,070;%%

\bibitem{BES}
N.~Beisert, B.~Eden and M.~Staudacher,
  ``Transcendentality and crossing,''
  J.\ Stat.\ Mech.\  {\bf 0701}, P021 (2007)
  [arXiv:hep-th/0610251].
  %%CITATION = JSTAT,0701,P021;%%

\bibitem{Belitsky:2006wg}
  A.~V.~Belitsky,
  ``Long-range SL(2) Baxter equation in N = 4 super-Yang-Mills theory,''
  Phys.\ Lett.\  B {\bf 643}, 354 (2006)
  [arXiv:hep-th/0609068].
  %%CITATION = PHLTA,B643,354;%%

\bibitem{AFS}
  G.~Arutyunov, S.~Frolov and M.~Staudacher,
  ``Bethe ansatz for quantum strings,''
  JHEP {\bf 0410}, 016 (2004)
  [arXiv:hep-th/0406256].
  %%CITATION = HEP-TH 0406256;%%

\bibitem{BK}
  N.~Beisert and T.~Klose,
  ``Long-range gl(n) integrable spin chains and plane-wave matrix theory,''
  J.\ Stat.\ Mech.\  {\bf 0607}, P006 (2006)
  [arXiv:hep-th/0510124].
  %%CITATION = HEP-TH 0510124;%%

\bibitem{HL}
  R.~Hernandez and E.~Lopez,
  ``Quantum corrections to the string Bethe ansatz,''
  JHEP {\bf 0607}, 004 (2006)
  [arXiv:hep-th/0603204].
  %%CITATION = HEP-TH 0603204;%%

\bibitem{Freyhult:2006vr}
  L.~Freyhult and C.~Kristjansen,
  ``A universality test of the quantum string Bethe ansatz,''
  Phys.\ Lett.\ B {\bf 638}, 258 (2006)
  [arXiv:hep-th/0604069].
  %%CITATION = HEP-TH 0604069;%%

\bibitem{HM}
  D.~M.~Hofman and J.~M.~Maldacena,
  ``Giant magnons,''
  J.\ Phys.\ A {\bf 39}, 13095 (2006)
  [arXiv:hep-th/0604135].
  %%CITATION = HEP-TH 0604135;%%

\bibitem{Janik}
  R.~A.~Janik,
  ``The AdS(5) x S**5 superstring worldsheet S-matrix and crossing symmetry,''
  Phys.\ Rev.\ D {\bf 73}, 086006 (2006)
  [arXiv:hep-th/0603038].
  %%CITATION = HEP-TH 0603038;%%

\bibitem{BCDKS}
Z.~Bern, M.~Czakon, L.~J.~Dixon, D.~A.~Kosower and V.~A.~Smirnov,
  ``The Four-Loop Planar Amplitude and Cusp Anomalous Dimension in Maximally
  Supersymmetric Yang-Mills Theory,''
  Phys.\ Rev.\  D {\bf 75}, 085010 (2007)
  [arXiv:hep-th/0610248].
  %%CITATION = PHRVA,D75,085010;%%

\bibitem{Cachazo:2006az}
  F.~Cachazo, M.~Spradlin and A.~Volovich,
  ``Four-Loop Cusp Anomalous Dimension From Obstructions,''
  Phys.\ Rev.\  D {\bf 75} (2007) 105011
  [arXiv:hep-th/0612309].
  %%CITATION = PHRVA,D75,105011;%%

\bibitem{BBKS}
  M.~K.~Benna, S.~Benvenuti, I.~R.~Klebanov and A.~Scardicchio,
  ``A test of the AdS/CFT correspondence using high-spin operators,''
  Phys.\ Rev.\ Lett.\  {\bf 98}, 131603 (2007)
  [arXiv:hep-th/0611135].
  %%CITATION = PRLTA,98,131603;%%

\bibitem{AABEK}
  L.~F.~Alday, G.~Arutyunov, M.~K.~Benna, B.~Eden and I.~R.~Klebanov,
  ``On the strong coupling scaling dimension of high spin operators,''
  JHEP {\bf 0704}, 082 (2007)
  [arXiv:hep-th/0702028].
  %%CITATION = JHEPA,0704,082;%%

\bibitem{Kostov}
  I.~Kostov, D.~Serban and D.~Volin,
  ``Strong coupling limit of Bethe ansatz equations,''
  Nucl.\ Phys.\  B {\bf 789}, 413 (2008)
  [arXiv:hep-th/0703031].
  %%CITATION = NUPHA,B789,413;%%

\bibitem{Basso}
  B.~Basso, G.~P.~Korchemsky and J.~Kotanski,
  ``Cusp anomalous dimension in maximally supersymmetric Yang-Mills theory at
  strong coupling,''
  arXiv:0708.3933 [hep-th].
  %%CITATION = ARXIV:0708.3933;%%

\bibitem{Kost}
  I.~Kostov, D.~Serban and D.~Volin,
  ``Functional BES equation,''
  arXiv:0801.2542 [hep-th].
  %%CITATION = ARXIV:0801.2542;%%

\bibitem{Roiban}
  R.~Roiban, A.~Tirziu and A.~A.~Tseytlin,
  ``Two-loop world-sheet corrections in $AdS_5$ x $S^5$ superstring,''
  JHEP {\bf 0707}, 056 (2007)
  [arXiv:0704.3638 [hep-th]];
  %%CITATION = JHEPA,0707,056;%%
R.~Roiban and A.~A.~Tseytlin,
  ``Strong-coupling expansion of cusp anomaly from quantum superstring,''
  JHEP {\bf 0711}, 016 (2007)
  [arXiv:0709.0681 [hep-th]].
  %%CITATION = JHEPA,0711,016;%%

\bibitem{gkp}
  S.~S.~Gubser, I.~R.~Klebanov and A.~W.~Peet,
  ``Entropy and Temperature of Black 3-Branes,''
  Phys.\ Rev.\ D {\bf 54}, 3915 (1996)
  [arXiv:hep-th/9602135];
  %%CITATION = HEP-TH 9602135;%%
I.~R.~Klebanov and A.~A.~Tseytlin,
  ``Entropy of Near-Extremal Black p-branes,''
  Nucl.\ Phys.\ B {\bf 475}, 164 (1996)
  [arXiv:hep-th/9604089].
  %%CITATION = HEP-TH 9604089;%%

\bibitem{GKT}
  S.~S.~Gubser, I.~R.~Klebanov and A.~A.~Tseytlin,
  ``Coupling constant dependence in the thermodynamics of N = 4  supersymmetric
  Yang-Mills theory,''
  Nucl.\ Phys.\ B {\bf 534}, 202 (1998)
  [arXiv:hep-th/9805156].
  %%CITATION = HEP-TH 9805156;%%

\bibitem{Foto}
  A.~Fotopoulos and T.~R.~Taylor,
  ``Comment on two-loop free energy in N = 4 supersymmetric Yang-Mills  theory
  at finite temperature,''
  Phys.\ Rev.\ D {\bf 59}, 061701 (1999)
  [arXiv:hep-th/9811224];
  %%CITATION = HEP-TH 9811224;%%
M.~A.~Vazquez-Mozo,
  ``A note on supersymmetric Yang-Mills thermodynamics,''
  Phys.\ Rev.\ D {\bf 60}, 106010 (1999)
  [arXiv:hep-th/9905030];
  %%CITATION = HEP-TH 9905030;%%
C.~J.~Kim and S.~J.~Rey,
  ``Thermodynamics of large-N super Yang-Mills theory and AdS/CFT
  correspondence,''
  Nucl.\ Phys.\ B {\bf 564}, 430 (2000)
  [arXiv:hep-th/9905205].
  %%CITATION = HEP-TH 9905205;%%

\bibitem{Karsch}
  F.~Karsch,
  ``Lattice QCD at high temperature and density,''
  Lect.\ Notes Phys.\  {\bf 583}, 209 (2002)
  [arXiv:hep-lat/0106019].
  %%CITATION = HEP-LAT 0106019;%%

\bibitem{Gavai}
  R.~V.~Gavai, S.~Gupta and S.~Mukherjee,
  ``A new method to determine the equation of state, specific heat, and speed
  of sound above and below the transition temperature in QCD,''
  arXiv:hep-lat/0506015.
  %%CITATION = HEP-LAT 0506015;%%

\bibitem{Bringoltz}
  B.~Bringoltz and M.~Teper,
``The pressure of the SU(N) lattice gauge theory at large-N,''
  Phys.\ Lett.\  B {\bf 628}, 113 (2005)
  [arXiv:hep-lat/0506034].
  %%CITATION = PHLTA,B628,113;%%

\bibitem{Policastro}
  G.~Policastro, D.~T.~Son and A.~O.~Starinets,
  ``The shear viscosity of strongly coupled N = 4 supersymmetric Yang-Mills
  plasma,''
  Phys.\ Rev.\ Lett.\  {\bf 87}, 081601 (2001)
  [arXiv:hep-th/0104066].
  %%CITATION = HEP-TH 0104066;%%

\bibitem{Huot}
  S.~C.~Huot, S.~Jeon and G.~D.~Moore,
  ``Shear viscosity in weakly coupled N = 4 super Yang-Mills theory  compared
  to QCD,''
  Phys.\ Rev.\ Lett.\  {\bf 98}, 172303 (2007)
  [arXiv:hep-ph/0608062].
  %%CITATION = PRLTA,98,172303;%%

\bibitem{Buchel}
  A.~Buchel, J.~T.~Liu and A.~O.~Starinets,
  ``Coupling constant dependence of the shear viscosity in N = 4 supersymmetric
  Yang-Mills theory,''
  Nucl.\ Phys.\ B {\bf 707}, 56 (2005)
  [arXiv:hep-th/0406264].
  %%CITATION = HEP-TH 0406264;%%

\bibitem{Kovtun}
  P.~Kovtun, D.~T.~Son and A.~O.~Starinets,
``Holography and hydrodynamics: Diffusion on stretched horizons,''
  JHEP {\bf 0310}, 064 (2003)
  [arXiv:hep-th/0309213];
  %%CITATION = HEP-TH 0309213;%%
  ``Viscosity in strongly interacting quantum field theories from black hole
  physics,''
  Phys.\ Rev.\ Lett.\  {\bf 94}, 111601 (2005)
  [arXiv:hep-th/0405231].
  %%CITATION = HEP-TH 0405231;%%

\bibitem{Cohen}
  T.~D.~Cohen,
  ``Is there a 'most perfect fluid' consistent with quantum field theory?,''
  Phys.\ Rev.\ Lett.\  {\bf 99}, 021602 (2007)
  [arXiv:hep-th/0702136].
  %%CITATION = PRLTA,99,021602;%%



\bibitem{Kats}
  Y.~Kats and P.~Petrov,
  ``Effect of curvature squared corrections in AdS on the viscosity of the dual
  gauge theory,''
  arXiv:0712.0743 [hep-th];
  %%CITATION = ARXIV:0712.0743;%%
M.~Brigante, H.~Liu, R.~C.~Myers, S.~Shenker and S.~Yaida,
  ``Viscosity Bound Violation in Higher Derivative Gravity,''
  arXiv:0712.0805 [hep-th].
  %%CITATION = ARXIV:0712.0805;%%

\bibitem{Teaney}
  D.~Teaney,
  ``Effect of shear viscosity on spectra, elliptic flow, and Hanbury
  Brown-Twiss radii,''
  Phys.\ Rev.\ C {\bf 68}, 034913 (2003).
  %%CITATION = PHRVA,C68,034913;%%

\bibitem{Hirano}
  T.~Hirano and M.~Gyulassy,
  ``Perfect fluidity of the quark gluon plasma core as seen through its
  dissipative hadronic corona,''
  arXiv:nucl-th/0506049.
  %%CITATION = NUCL-TH 0506049;%%

\bibitem{Romatschke}
  P.~Romatschke and U.~Romatschke,
  ``Viscosity Information from Relativistic Nuclear Collisions: How Perfect is
  the Fluid Observed at RHIC?,''
  Phys.\ Rev.\ Lett.\  {\bf 99}, 172301 (2007)
  [arXiv:0706.1522 [nucl-th]].
  %%CITATION = PRLTA,99,172301;%%

\bibitem{Meyer}
  H.~B.~Meyer,
  ``A calculation of the shear viscosity in SU(3) gluodynamics,''
  Phys.\ Rev.\  D {\bf 76}, 101701 (2007)
  [arXiv:0704.1801 [hep-lat]].
  %%CITATION = PHRVA,D76,101701;%%

\bibitem{Gyulassy}
  M.~Gyulassy and L.~McLerran,
  ``New forms of QCD matter discovered at RHIC,''
  Nucl.\ Phys.\ A {\bf 750}, 30 (2005)
  [arXiv:nucl-th/0405013].
  %%CITATION = NUCL-TH 0405013;%%

\bibitem{Shuryak}
  E.~V.~Shuryak,
  ``What RHIC experiments and theory tell us about properties of quark-gluon
  plasma?,''
  Nucl.\ Phys.\ A {\bf 750}, 64 (2005)
  [arXiv:hep-ph/0405066].
  %%CITATION = HEP-PH 0405066;%%

\bibitem{Pisarsky}
R. Pisarsky,
talk available at http://quark.phy.bnl.gov/~pisarski/talks/unicorn.pdf . 

\bibitem{ks}
S.~Kachru and E.~Silverstein,
``4d conformal theories and strings on orbifolds,''
Phys.\ Rev.\ Lett.\  {\bf 80}, 4855 (1998)
[arXiv:hep-th/9802183]; \\
%%CITATION = HEP-TH 9802183;%%
A.~E.~Lawrence, N.~Nekrasov and C.~Vafa,
``On conformal field theories in four dimensions,''
Nucl.\ Phys.\ B {\bf 533}, 199 (1998)
[arXiv:hep-th/9803015].
%%CITATION = HEP-TH 9803015;%%

\bibitem{Kehag}
A.~Kehagias,
``New type IIB vacua and their F-theory interpretation,''
Phys.\ Lett.\ B {\bf 435}, 337 (1998)
[arXiv:hep-th/9805131].
%%CITATION = HEP-TH 9805131;%%

\bibitem{KW}
I.~R.~Klebanov and E.~Witten,
``Superconformal field theory on threebranes at a Calabi-Yau  singularity,''
Nucl.\ Phys.\ B {\bf 536}, 199 (1998)
[arXiv:hep-th/9807080].
%%CITATION = HEP-TH 9807080;%%

\bibitem{Morrison}
D.~Morrison and R.~Plesser,
``Non-Spherical Horizons, I,''
Adv. Theor. Math. Phys. {\bf 3} (1999) 1,  [arXiv:hep-th/9810201].
%%CITATION = HEP-TH 9810201;%%

\bibitem{Herzog}
C.~P.~Herzog, I.~R.~Klebanov and P.~Ouyang,
``D-branes on the conifold and N = 1 gauge / gravity dualities,''
arXiv:hep-th/0205100.
%%CITATION = HEP-TH 0205100;%%

%\cite{Candelas:1989js}
\bibitem{Candelas}
  P.~Candelas and X.~C.~de la Ossa,
  ``Comments on conifolds,''
  Nucl.\ Phys.\ B {\bf 342}, 246 (1990).
  %%CITATION = NUPHA,B342,246;%%

\bibitem{Romans}
L.~Romans, ``New compactifications of chiral $N=2$, $d=10$
supergravity,''  Phys. Lett. {\bf B153} (1985) 392.

\bibitem{Klebanov:1999tb}
  I.~R.~Klebanov and E.~Witten,
  ``AdS/CFT correspondence and symmetry breaking,''
  Nucl.\ Phys.\  B {\bf 556} (1999) 89
  [arXiv:hep-th/9905104].
%%CITATION = NUPHA,B556,89;%%

\bibitem{Klebanov:2007us}
  I.~R.~Klebanov and A.~Murugan,
  ``Gauge / gravity duality and warped resolved conifold,''
  JHEP {\bf 0703} (2007) 042
  [arXiv:hep-th/0701064].
%%CITATION = JHEPA,0703,042;%%

\bibitem{Pando:2000sq}
  L.~A.~Pando Zayas and A.~A.~Tseytlin,
  ``3-branes on resolved conifold,''
  JHEP {\bf 0011}, 028 (2000)
  [arXiv:hep-th/0010088].
  %%CITATION = JHEPA,0011,028;%%

\bibitem{KMRW}
  I.~R.~Klebanov, A.~Murugan, D.~Rodriguez-Gomez and J.~Ward,
  ``Goldstone Bosons and Global Strings in a Warped Resolved Conifold,''
  arXiv:0712.2224 [hep-th].
  %%CITATION = ARXIV:0712.2224;%%

\bibitem{GK}
  S.~S.~Gubser and I.~R.~Klebanov,
  ``Baryons and domain walls in an N = 1 superconformal gauge theory,''
  Phys.\ Rev.\ D {\bf 58}, 125025 (1998)
  [arXiv:hep-th/9808075].
  %%CITATION = HEP-TH 9808075;%%

\bibitem{KW2}
  I.~R.~Klebanov and E.~Witten,
  ``AdS/CFT correspondence and symmetry breaking,''
  Nucl.\ Phys.\ B {\bf 556}, 89 (1999)
  [arXiv:hep-th/9905104].
  %%CITATION = HEP-TH 9905104;%%

\bibitem{Georgi:2007ek}
  H.~Georgi,
  ``Unparticle Physics,''
  Phys.\ Rev.\ Lett.\  {\bf 98} (2007) 221601
  [arXiv:hep-ph/0703260].
  %%CITATION = PRLTA,98,221601;%%

\bibitem{KS}
I.~R.~Klebanov and M.~J.~Strassler,
``Supergravity and a confining gauge theory: Duality cascades and
chiSB-resolution of naked singularities,''
JHEP {\bf 0008}, 052 (2000)
[arXiv:hep-th/0007191].
%%CITATION = HEP-TH 0007191;%%

\bibitem{Seiberg}
  N.~Seiberg,
  ``Electric - magnetic duality in supersymmetric nonAbelian gauge theories,''
  Nucl.\ Phys.\ B {\bf 435}, 129 (1995)
  [arXiv:hep-th/9411149].
  %%CITATION = HEP-TH 9411149;%%

\bibitem{Strassler}
  M.~J.~Strassler,
  ``The duality cascade,''
  arXiv:hep-th/0505153.
  %%CITATION = HEP-TH 0505153;%%

\bibitem{GHK}
  S.~S.~Gubser, C.~P.~Herzog and I.~R.~Klebanov,
  ``Symmetry breaking and axionic strings in the warped deformed conifold,''
  JHEP {\bf 0409}, 036 (2004)
  [arXiv:hep-th/0405282];
  %%CITATION = HEP-TH 0405282;%%
``Variations on the warped deformed conifold,''
  Comptes Rendus Physique {\bf 5}, 1031 (2004)
  [arXiv:hep-th/0409186].
  %%CITATION = CRPOB,5,1031;%%

\bibitem{PT}
G.~Papadopoulos and A.~A.~Tseytlin,
``Complex geometry of conifolds and 5-brane wrapped on 2-sphere,''
Class.\ Quant.\ Grav.\  {\bf 18}, 1333 (2001)
[arXiv:hep-th/0012034].
%%CITATION = HEP-TH 0012034;%%

\bibitem{Ceres}
A.~Ceresole, G.~Dall'Agata, R.~D'Auria and S.~Ferrara,
``Spectrum of type IIB supergravity on AdS(5) x T(11): Predictions on N  = 1
SCFT's,''
Phys.\ Rev.\ D {\bf 61}, 066001 (2000)
[arXiv:hep-th/9905226].
%%CITATION = HEP-TH 9905226;%%

\bibitem{Aharony}
O.~Aharony,
``A note on the holographic interpretation of string theory backgrounds  with
varying flux,''
JHEP {\bf 0103}, 012 (2001)
[arXiv:hep-th/0101013].
%%CITATION = HEP-TH 0101013;%%


\bibitem{Seiberg:1994bz}
  N.~Seiberg,
  ``Exact Results On The Space Of Vacua Of Four-Dimensional Susy Gauge
  Theories,''
  Phys.\ Rev.\  D {\bf 49}, 6857 (1994)
  [arXiv:hep-th/9402044].
  %%CITATION = PHRVA,D49,6857;%%

\bibitem{Krasnitz}
  M.~Krasnitz,
  ``A two point function in a cascading N = 1 gauge theory from
  supergravity,''
  arXiv:hep-th/0011179;
  %%CITATION = HEP-TH/0011179;%%
  ``Correlation functions in a cascading N = 1 gauge theory from
  supergravity,''
  JHEP {\bf 0212}, 048 (2002)
  [arXiv:hep-th/0209163].
  %%CITATION = JHEPA,0212,048;%%


\bibitem{Caceres}
  E.~Caceres,
  ``A Brief Review Of Glueball Masses From Gauge / Gravity Duality,''
  J.\ Phys.\ Conf.\ Ser.\  {\bf 24}, 111 (2005).
  %%CITATION = 00462,24,111;%%


\bibitem{BHM}
M.~Berg, M.~Haack and W.~Mueck,
  ``Bulk dynamics in confining gauge theories,''
  Nucl.\ Phys.\  B {\bf 736}, 82 (2006)
  [arXiv:hep-th/0507285];
  %%CITATION = NUPHA,B736,82;%%
``Glueballs vs. gluinoballs: Fluctuation spectra in non-AdS/non-CFT,''
  Nucl.\ Phys.\  B {\bf 789}, 1 (2008)
  [arXiv:hep-th/0612224].
  %%CITATION = NUPHA,B789,1;%%

\bibitem{Karch}
  A.~Karch, E.~Katz, D.~T.~Son and M.~A.~Stephanov,
  ``Linear confinement and AdS/QCD,''
  Phys.\ Rev.\  D {\bf 74}, 015005 (2006)
  [arXiv:hep-ph/0602229].
  %%CITATION = PHRVA,D74,015005;%%

\bibitem{BDKS}
  M.~K.~Benna, A.~Dymarsky, I.~R.~Klebanov and A.~Solovyov,
  ``On Normal Modes of a Warped Throat,''
  arXiv:0712.4404 [hep-th].
  %%CITATION = ARXIV:0712.4404;%%

\bibitem{Butti}
  A.~Butti, M.~Grana, R.~Minasian, M.~Petrini and A.~Zaffaroni,
  ``The baryonic branch of Klebanov-Strassler solution: A supersymmetric family
  of SU(3) structure backgrounds,''
  JHEP {\bf 0503}, 069 (2005)
  [arXiv:hep-th/0412187].
  %%CITATION = HEP-TH 0412187;%%

\bibitem{DKS}
  A.~Dymarsky, I.~R.~Klebanov and N.~Seiberg,
  ``On the moduli space of the cascading SU(M+p) x SU(p) gauge theory,''
  JHEP {\bf 0601}, 155 (2006)
  [arXiv:hep-th/0511254].
  %%CITATION = JHEPA,0601,155;%%

\bibitem{KT}
I.~R.~Klebanov and A.~A.~Tseytlin,
``Gravity duals of supersymmetric SU(N) x SU(N+M) gauge theories,''
Nucl.\ Phys.\ B {\bf 578}, 123 (2000)
[arXiv:hep-th/0002159].
%%CITATION = HEP-TH 0002159;%%

\bibitem{CVMN}
A.~H.~Chamseddine and M.~S.~Volkov,
  ``Non-Abelian BPS monopoles in N = 4 gauged supergravity,''
  Phys.\ Rev.\ Lett.\  {\bf 79}, 3343 (1997)
  [arXiv:hep-th/9707176];
  %%CITATION = PRLTA,79,3343;%%
  ``Non-Abelian solitons in N = 4 gauged supergravity and leading order  string
  theory,''
  Phys.\ Rev.\  D {\bf 57}, 6242 (1998)
  [arXiv:hep-th/9711181];
  %%CITATION = PHRVA,D57,6242;%%
  J.~M.~Maldacena and C.~Nunez,
  ``Towards the large N limit of pure N = 1 super Yang Mills,''
  Phys.\ Rev.\ Lett.\  {\bf 86}, 588 (2001)
  [arXiv:hep-th/0008001].
  %%CITATION = HEP-TH 0008001;%%

\bibitem{BDK}
  M.~K.~Benna, A.~Dymarsky and I.~R.~Klebanov,
  ``Baryonic condensates on the conifold,''
  JHEP {\bf 0708}, 034 (2007)
  [arXiv:hep-th/0612136].
  %%CITATION = JHEPA,0708,034;%%

\bibitem{Aharony:2005zr}
  O.~Aharony, A.~Buchel and A.~Yarom,
  ``Holographic renormalization of cascading gauge theories,''
  Phys.\ Rev.\ D {\bf 72}, 066003 (2005)
  [arXiv:hep-th/0506002].
  %%CITATION = HEP-TH 0506002;%%

\bibitem{Aharony:2006ce}
  O.~Aharony, A.~Buchel and A.~Yarom,
  ``Short distance properties of cascading gauge theories,''
JHEP {\bf 0611}, 069 (2006)
  [arXiv:hep-th/0608209].
  %%CITATION = JHEPA,0611,069;%%

\bibitem{Marino:1999af}
  M.~Marino, R.~Minasian, G.~W.~Moore and A.~Strominger,
  ``Nonlinear instantons from supersymmetric p-branes,''
  JHEP {\bf 0001}, 005 (2000)
  [arXiv:hep-th/9911206].
  %%CITATION = HEP-TH 9911206;%%

\bibitem{Inflation}
  A.~H.~Guth,
  ``The Inflationary Universe: A Possible Solution To The Horizon And Flatness
  Problems,''
  Phys.\ Rev.\ D {\bf 23}, 347 (1981);
 A.~D.~Linde,
  ``A New Inflationary Universe Scenario: A Possible Solution Of The Horizon,
  Flatness, Homogeneity, Isotropy And Primordial Monopole Problems,''
  Phys.\ Lett.\ B {\bf 108}, 389 (1982);
  A.~Albrecht and P.~J.~Steinhardt,
  ``Cosmology For Grand Unified Theories With Radiatively Induced Symmetry
  Breaking,''
  Phys.\ Rev.\ Lett.\  {\bf 48}, 1220 (1982).

\bibitem{reviews}
%\cite{Linde:2005dd}
%\bibitem{LindeReview}
  A.~Linde,
  ``Inflation and string cosmology,''
  eConf {\bf C040802}, L024 (2004)
  [J.\ Phys.\ Conf.\ Ser.\  {\bf 24}, 151 (2005\ PTPSA,163,295-322.2006)]
  [arXiv:hep-th/0503195];
  %%CITATION = HEP-TH 0503195;%%
%\cite{HenryTye:2006uv}
%\bibitem{TyeReview}
  S.~H.~Henry Tye,
  ``Brane inflation: String theory viewed from the cosmos,''
  arXiv:hep-th/0610221;
  %%CITATION = HEP-TH 0610221;%%
 %\cite{Cline:2006hu}
%\bibitem{ClineReview}
  J.~M.~Cline,
  ``String cosmology,''
  arXiv:hep-th/0612129;
  %%CITATION = HEP-TH 0612129;%%
%\cite{Kallosh:2007ig}
%\bibitem{KalloshReview}
  R.~Kallosh,
  ``On Inflation in String Theory,''
  arXiv:hep-th/0702059.
%\bibitem{McAllister:2007bg}
  L.~McAllister and E.~Silverstein,
  ``String Cosmology: A Review,''
  arXiv:0710.2951 [hep-th].
  %%CITATION = ARXIV:0710.2951;%%



\bibitem{Dvali}
  G.~R.~Dvali and S.~H.~H.~Tye,
  ``Brane inflation,''
  Phys.\ Lett.\ B {\bf 450}, 72 (1999) [arXiv:hep-ph/9812483].
  %%CITATION = HEP-PH 9812483;%%


\bibitem{KKLMMT}
  S.~Kachru, R.~Kallosh, A.~Linde, J.~M.~Maldacena, L.~P.~McAllister and S.~P.~Trivedi,
  ``Towards inflation in string theory,''
  JCAP {\bf 0310}, 013 (2003)
  [arXiv:hep-th/0308055].
  %%CITATION = JCAPA,0310,013;%%


\bibitem{Hebecker}
  A.~Hebecker and J.~March-Russell,
  ``The ubiquitous throat,''
  Nucl.\ Phys.\  B {\bf 781}, 99 (2007)
  [arXiv:hep-th/0607120].
  %%CITATION = NUPHA,B781,99;%%

\bibitem{KKLT}
  S.~Kachru, R.~Kallosh, A.~Linde and S.~P.~Trivedi,
  ``De Sitter vacua in string theory,''
  Phys.\ Rev.\  D {\bf 68}, 046005 (2003)
  [arXiv:hep-th/0301240].
  %%CITATION = PHRVA,D68,046005;%%

\bibitem{BDKMMM}
  D.~Baumann, A.~Dymarsky, I.~R.~Klebanov, J.~M.~Maldacena, L.~P.~McAllister and A.~Murugan,
  ``On D3-brane potentials in compactifications with fluxes and wrapped
  D-branes,''
  JHEP {\bf 0611}, 031 (2006)
  [arXiv:hep-th/0607050].
  %%CITATION = JHEPA,0611,031;%%

\bibitem{Baumann:2007ah}
  D.~Baumann, A.~Dymarsky, I.~R.~Klebanov and L.~McAllister,
  ``Towards an Explicit Model of D-brane Inflation,''
  JCAP {\bf 0801}, 024 (2008)
  [arXiv:0706.0360 [hep-th]].
  %%CITATION = JCAPA,0801,024;%%
D.~Baumann, A.~Dymarsky, I.~R.~Klebanov, L.~McAllister and P.~J.~Steinhardt,
  ``A Delicate Universe: Compactification Obstacles to D-Brane Inflation,''
  Phys.\ Rev.\ Lett.\  {\bf 99}, 141601 (2007)
  [arXiv:0705.3837 [hep-th]].
  %%CITATION = PRLTA,99,141601;%%


\bibitem{GKP}
  S.~B.~Giddings, S.~Kachru and J.~Polchinski,
  ``Hierarchies from fluxes in string compactifications,''
  Phys.\ Rev.\  D {\bf 66}, 106006 (2002)
  [arXiv:hep-th/0105097].
  %%CITATION = PHRVA,D66,106006;%%


\bibitem{JPolchinski}
  J.~Polchinski,
  ``Introduction to cosmic F- and D-strings,''
  arXiv:hep-th/0412244.
  %%CITATION = HEP-TH/0412244;%%

\bibitem{Copeland}
  E.~J.~Copeland, R.~C.~Myers and J.~Polchinski,
  ``Cosmic F- and D-strings,''
  JHEP {\bf 0406}, 013 (2004)
  [arXiv:hep-th/0312067].
  %%CITATION = JHEPA,0406,013;%%

\bibitem{HK}
  C.~P.~Herzog and I.~R.~Klebanov,
  ``On string tensions in supersymmetric SU(M) gauge theory,''
  Phys.\ Lett.\  B {\bf 526}, 388 (2002)
  [arXiv:hep-th/0111078].
  %%CITATION = PHLTA,B526,388;%%


\bibitem{Firouzjahi}
  H.~Firouzjahi, L.~Leblond and S.~H.~Henry Tye,
  ``The (p,q) string tension in a warped deformed conifold,''
  JHEP {\bf 0605}, 047 (2006)
  [arXiv:hep-th/0603161].
  %%CITATION = JHEPA,0605,047;%%

\bibitem{Sarangi}
  S.~Sarangi and S.~H.~H.~Tye,
  ``Cosmic string production towards the end of brane inflation,''
  Phys.\ Lett.\  B {\bf 536}, 185 (2002)
  [arXiv:hep-th/0204074].
  %%CITATION = PHLTA,B536,185;%%

\bibitem{DeWG}
  O.~DeWolfe and S.~B.~Giddings,
  ``Scales and hierarchies in warped compactifications and brane worlds,''
  Phys.\ Rev.\ D {\bf 67}, 066008 (2003)
  [arXiv:hep-th/0208123].

\bibitem{GM}
  S.~B.~Giddings and A.~Maharana,
  ``Dynamics of warped compactifications and the shape of the warped
  landscape,''
  Phys.\ Rev.\  D {\bf 73}, 126003 (2006)
  [arXiv:hep-th/0507158].
  %%CITATION = PHRVA,D73,126003;%%


\bibitem{Krause}
  A.~Krause and E.~Pajer,
  ``Chasing Brane Inflation in String-Theory,''
  arXiv:0705.4682 [hep-th].
  %%CITATION = ARXIV:0705.4682;%%

\bibitem{Holman}
  R.~Holman, P.~Ramond and G.~G.~Ross,
  ``Supersymmetric Inflationary Cosmology,''
  Phys.\ Lett.\  B {\bf 137}, 343 (1984).
  %%CITATION = PHLTA,B137,343;%%

\bibitem{MSSM}
  R.~Allahverdi, K.~Enqvist, J.~Garcia-Bellido and A.~Mazumdar,
  ``Gauge invariant MSSM inflaton,''
  Phys.\ Rev.\ Lett.\  {\bf 97}, 191304 (2006)
  [arXiv:hep-ph/0605035];
 R.~Allahverdi, K.~Enqvist, J.~Garcia-Bellido, A.~Jokinen and A.~Mazumdar,
 ``MSSM flat direction inflation: Slow roll, stability, fine tunning and
 reheating,''
 JCAP {\bf 0706}, 019 (2007)
 [arXiv:hep-ph/0610134];
 J.~C.~Bueno Sanchez, K.~Dimopoulos and D.~H.~Lyth,
``A-term inflation and the MSSM,''
JCAP {\bf 0701}, 015 (2007)
[arXiv:hep-ph/0608299].

\bibitem{Itzhaki}
  N.~Itzhaki and E.~D.~Kovetz,
  ``Inflection Point Inflation and Time Dependent Potentials in String
  Theory,''
  arXiv:0708.2798 [hep-th]; see also N. Itzhaki's talk at the
Conference ``String Theory: Achievements and Perspectives,''
Jerusalem and Tel-Aviv April 2007,  http://stringfest.tau.ac.il
  %%CITATION = ARXIV:0708.2798;%%

\bibitem{Linde}
  A.~Linde and A.~Westphal,
  ``Accidental Inflation in String Theory,''
  arXiv:0712.1610 [hep-th].
  %%CITATION = ARXIV:0712.1610;%%

\bibitem{Underwood}
  B.~Underwood,
  ``Brane Inflation is Attractive,''
  arXiv:0802.2117 [hep-th].
  %%CITATION = ARXIV:0802.2117;%%


\end{thebibliography}
\end{document}